\documentclass[12pt]{article}

\usepackage[utf8]{inputenc}
\usepackage[T1]{fontenc}
\usepackage{lmodern}
\usepackage{newunicodechar}
\newunicodechar{√}{\ensuremath{\sqrt{}}}
\usepackage{geometry}
\usepackage{hyperref}
\usepackage{graphicx}
\usepackage{amsmath}
\usepackage{authblk}
\usepackage[numbers,sort&compress]{natbib}
\usepackage{color}

\usepackage{tabularx}

\title{Cluster Haptic Texture Dataset:\\Haptic Texture Dataset with Varied Velocity–Direction Sliding Contacts}

\author[1,2]{Michikuni~Eguchi\thanks{m.eguchi@cluster.mu}}
\author[1]{Tomohiro~Hayase\thanks{t.hayase@cluster.mu}}
\author[1]{Yuichi~Hiroi\thanks{y.hiroi@cluster.mu}}
\author[3,1]{Takefumi~Hiraki\thanks{Corresponding author: t.hiraki@cluster.mu}}

\affil[1]{Cluster Metaverse Lab, 8-9-5 Nishigotanda, Shinagawa, Tokyo, Japan}
\affil[2]{Graduate School of Comprehensive Human Sciences, University of Tsukuba, 1-2 Kasuga, Tsukuba, Ibaraki, Japan}
\affil[3]{Institute of Library, Information and Media Science, University of Tsukuba, 1-2 Kasuga, Tsukuba, Ibaraki Japan}

\date{}

\begin{document}
\maketitle

% ---------- Abstract ----------

\begin{abstract}
Haptic sciences and technologies benefit greatly from comprehensive datasets that capture tactile stimuli under controlled, systematic conditions. However, existing haptic datasets collect data through uncontrolled exploration, which hinders the systematic analysis of how motion parameters (e.g., motion direction and velocity) influence tactile perception. This paper introduces Cluster Haptic Texture Dataset, a multimodal dataset recorded using a 3-axis machine with an rubber tip as artificial finger to precisely control sliding velocity and direction. The dataset encompasses 118 textured surfaces across 9 material categories, with recordings at 5 velocity levels (20-60 mm/s) and 8 directions. Each surface was tested under 160 conditions, yielding 18,880 synchronized recordings of audio, acceleration, force, position, and visual data. Validation using convolutional neural networks demonstrates classification accuracies of 96\% for texture recognition, 88.76\% for velocity estimation, and 78.79\% for direction estimation, confirming the dataset's utility for machine learning applications. This resource enables research in haptic rendering, texture recognition algorithms, and human tactile perception mechanisms, supporting the development of realistic haptic interfaces for virtual reality and robotic applications.
\end{abstract}

% =============================================================
%                        MAIN SECTIONS
% =============================================================

\section*{Background \& Summary}

% 1. マルチモーダル知覚
Human perception is known to be multisensory, and we perceive sensations by integrating different sources of information accessed through different sensory modalities in the brain.
Within this framework, kinesthetic perception, the sense of limb movement, and tactile perception are particularly important for recognizing objects and their surface properties. When we investigate an object with the hand, kinesthetic perception indicates how we are moving relative to the object, whereas tactile perception conveys the surface stimuli produced by that movement. Integrating these submodalities enables humans to infer object properties efficiently~\cite{Lederman2009-tf, ryan2021interaction, roberts2024visual}.
This principle also informs the design of human-machine interfaces. Systems that detect the user's movements and synchronize tactile, visual, and auditory feedback to those movements more closely replicate natural perception, thereby enhancing user immersion and realism~\cite{El_Saddik2007-wl, Bau2010-nk, Miyatake2023-bo, Ito2022-vr}.

% 2. 触覚データセットの必要性
In haptics research, large-scale data resources are indispensable for both fundamental and applied progress. Many studies on texture recognition and haptic rendering collect extensive sensor recordings from diverse real‑world tactile stimuli and leverage these data for tasks such as analysis~\cite{Wiertlewski2011-es, Platkiewicz2014-th}, model training~\cite{8793967, 10.1145/3586183.3606764, huang2021texture}, and haptic playback~\cite{Culbertson2013-ut, 10.1109/TOH.2024.3521418}.
Furthermore, experience in related fields shows that releasing large‑scale datasets openly accelerates research progress. For example, ImageNet~\cite{Deng2009-uz} serves as a benchmark in computer vision, while AudioSet~\cite{7952261} plays a similar role in audio analysis. An open haptic dataset could similarly accelerate progress.
However, for haptic datasets, it is not sufficient to release only tactile signals such as forces or vibrations. Because tactile perception is inherently coupled with cues from other modalities, such as visual and auditory, it is essential to collect multimodal recordings that synchronize haptic, visual, and auditory streams.

% 3.触覚データセットの現状
In response to this demand for haptic datasets, several groups have released open‑source haptic datasets that include various sensory modalities such as visual, audio, acceleration, and force~\cite{Culbertson2014-go, Zheng2016-ge, Strese2014-ye, strese2016multimodal, 7989057}.
In addition, the interaction between the probe and texture is a crucial element in haptic texture data. Several studies have focused on developing haptic datasets that specifically examine these probe-texture interactions, considering factors such as probe shape~\cite{10.1109/TOH.2024.3356609} and deformation~\cite{Jiao2019-hd, devillard2025tactile, 8794285}.
We summarize the key features of each haptic dataset in Table~\ref{tab:related_datasets}.

\begin{figure}[t]
    \centering
    \includegraphics[width=\columnwidth]{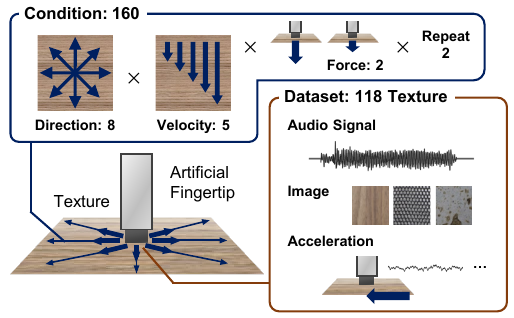}
    \caption{Concept of our Cluster Haptic Texture Dataset. We present a novel approach to haptic texture data collection by integrating controlled sliding velocities and directions using a rubber tip as an artificial finger, thus enhancing the conventional haptic dataset framework with precise motion parameters.}
    \label{fig:concept}
\end{figure}

% 4. 触覚データセットの課題
However, many existing multimodal haptic datasets omit systematic control of motion parameters such as sliding velocity and direction, because they are collected under free‑exploration conditions~\cite{Culbertson2014-go, Zheng2016-ge, Strese2014-ye, strese2016multimodal, 7989057, 10.1109/TOH.2024.3356609}.
These parameters strongly influence haptic perception, as variations in velocity and direction change the spatiotemporal pattern of skin deformation, and their systematic collection is crucial for subsequent utilization of haptic data.
There have been attempts to control motion, either by giving participants GUI‑based instructions~\cite{Jiao2019-hd, devillard2025tactile} or by employing machine‑controlled sliding~\cite{8794285}, but the range of motion variations in these studies remains very limited.

% 5. 提案内容
To address this gap, we propose the Cluster Haptic Texture Dataset, a comprehensive multimodal dataset under controlled sliding conditions with five velocity levels, eight motion directions. 
We recorded multimodal signals including applied load, contact-generated sound, and acceleration at the fingertip using a three-axis machine that enables precise motion control.
We collected data from 118 textured surfaces using a rubber tip as an artificial fingertip that mimics the mechanical properties of human skin.
The overall concept of our dataset is illustrated in Fig.~\ref{fig:concept}.

% 6. 論文内容、貢献
In this paper, we conducted texture classification experiments comparing different sensor modalities commonly used in previous studies, as well as their combinations with probe motion information (velocity and direction) to validate the effectiveness of our dataset.
In addition, we compared the accuracy of velocity and direction classification for each surface to evaluate whether sufficient velocity- and direction-related features are included for each texture material.
In summary, the contributions of this paper are as follows:

\begin{itemize}
    \item We propose the Cluster Haptic Texture Dataset, a multimodal dataset with controlled sliding parameters, containing visual, force, audio, and acceleration data from 118 textured surfaces.
    
    \item We developed a measurement system that enables systematic haptic data collection with precisely controlled sliding velocities and directions.
    
    \item We validate the effectiveness of our dataset through texture, sliding velocity, and direction classification experiments using the collected haptic data.
\end{itemize}

Our dataset, which incorporates a wide variety of explicitly controlled motion patterns, has strong potential for some haptic feedback methods and robot control methods. As a haptic feedback method, it enables the generation of realistic tactile stimuli that reflect diverse motion conditions~\cite{Cai2022-tg, Song2023-yv} and supports the design of human–machine interfaces such as smartphones and VR devices that provide richer and more immersive user experiences~\cite{El_Saddik2007-wl, Bau2010-nk, Miyatake2023-bo, Ito2022-vr}. In robotics, the dataset contributes to more precise modeling of touch and promotes the utilization of multimodal information~\cite{liu2024maniwav, li2023see}.

\begin{table*}[t]
\centering
\caption{Comparison of existing haptic texture datasets. This table provides a comparative overview of datasets, summarizing the number of materials, types of modalities, probe material, motion conditions, and motion varieties.}
\label{tab:related_datasets}
\scalebox{0.75}{
\begin{tabularx}{1.3\textwidth}{|l||l|X|l|X|X|}
\hline
\textbf{Name} & \textbf{Materials} & \textbf{Modalities} & \textbf{Probe} & \textbf{Motion conditions} & \textbf{Motion varieties} \\
\hline
HaTT~\cite{Culbertson2014-go} & 100 & Acceleration, force, position and image & Metal tip & Free-exploration slide & velocity, \newline direction, force \\
\hline
LMT~\cite{Zheng2016-ge, Strese2014-ye, strese2016multimodal} & 69 & Acceleration, audio, force, position and image & Metal tip & Free-exploration slide & velocity, \newline direction, force \\
\hline
Proton 2~\cite{7989057} & 5 & Acceleration, force, position and image & Metal tip & Free-exploration slide & velocity, \newline direction, force \\
\hline
MPI-10~\cite{10.1109/TOH.2024.3356609} & 10 & Acceleration, audio, force, position and image & Metal tip & Free-exploration slide & velocity, \newline direction, force \\
\hline
HapTex~\cite{Jiao2019-hd} & 120 & Force, friction coef., velocity, position and image & Human finger & Human-instruction-followed slide & none \newline (single condition) \\
\hline
\cite{devillard2025tactile} & 10 & Acceleration, audio, force, position and image & Human finger & Human-instruction-followed slide & direction \\
\hline
PFN-VT~\cite{8794285} & 25 & Force and image & Silicon rubber & Machine-controlled slide & none \newline (single condition) \\
\hline\hline
Ours & 118 & Acceleration, Audio, force, friction coef., position and image & Urethane rubber & Machine-controlled slide & velocity, \newline direction, force \\
\hline
\end{tabularx}
}
\end{table*}

\section*{Methods}
This section presents information about the measurement equipment, the procedures for acquiring texture data, and the texture surface materials in the Cluster Haptic Texture Dataset.

\subsection*{Hardware Setup}
\begin{figure}[t]
    \centering
    \includegraphics[width=\columnwidth]{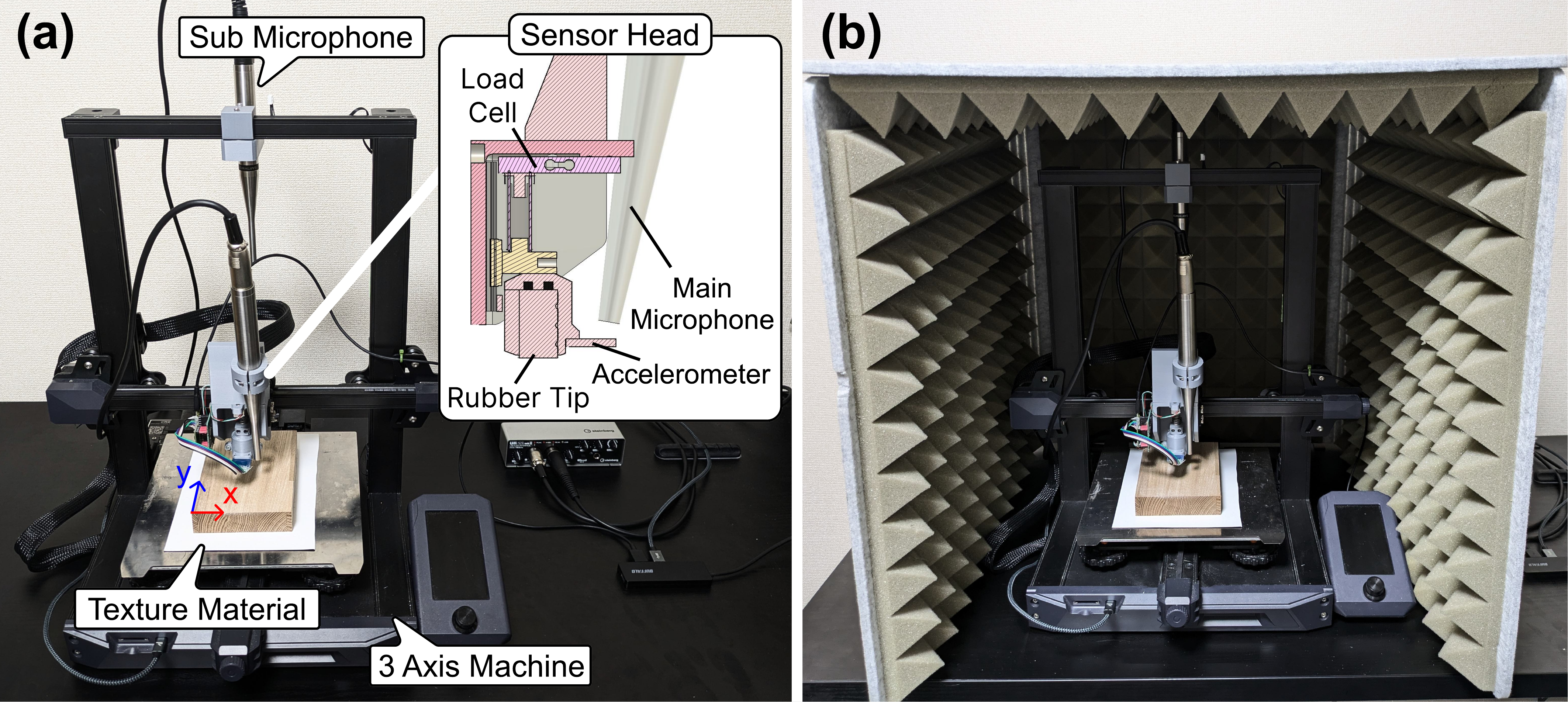}
    \caption{Haptic recording system overview: (a) Hardware setup for texture data measurement system. (b) The equipment is placed in a soundproof environment during data recording.}
    \label{fig:hardware}
\end{figure}
To acquire the sliding interaction information of the textured surface at different and controlled velocities and directions, we developed a special measuring device based on a 3-axis machine to construct a texture dataset using this information.
Figure~\ref{fig:hardware}(a) shows the configuration of the measurement system.
The system consists of a 3-axis machine, an artificial urethane rubber fingertip attached to the 3-axis machine as an end-effector, a microphone, and a load cell and accelerometer attached to the artificial finger.
The 3-axis machine and all sensors were connected to a single PC (ThinkPad X13 Gen 2, Lenovo), and the measurement information was recorded along with time stamps, with a PC clock accuracy of ±1 µs.
To capture the subtle sounds generated when the artificial fingertip contacts the textured surfaces, we performed all measurements in a soundproof enclosure, as shown in Figure~\ref{fig:hardware}(b).
The details of the components are described below.

\subsubsection*{3-axis Machine}
We used a modified 3D printer (Ender-3 S1, Creality) as the 3-axis machine, where the filament ejection unit was removed to accommodate the end-effector.
The 3D printer's modeling bed sheet was removed to expose the metal surface.
This is because we used a magnetic sheet to attach the textured material to this bed.
We controlled the operation of the 3-axis machine by sending G-code commands from a PC connected to the 3-axis machine to specify movement positions and velocities.
We also received position information from the 3-axis machine at 100 Hz and recorded it as part of the dataset.
In terms of accuracy, the machine has a positional precision of ±0.05 mm and a velocity control accuracy of ±0.2 mm/s. In addition, when integrated with a load cell, it enables force control with an accuracy of approximately ±0.05 N.

\subsubsection*{Rubber Tip}
We used a cylindrical urethane rubber tip (SCXN15-25, MISUMI) as the artificial fingertip attached to the end-effector.
The hardness of the rubber tip was Shore A15, which was determined based on the results of a previous study~\cite{Kuramitsu2013-os}.
The diameter of the cylindrical rubber tip was 15 mm, and this value was determined based on the average index finger contact area (Ellipse with a major axis radius of 6.8 mm and minor axis radius of 5.7 mm, area of about 122 mm$^2$) in the AIST hand dimension dataset~\cite{Kouchi2005-pq, Artificial_Intelligence_Research_Center_AIST_undated-jc}, ensuring that the cross-sectional area of the cylinder was close to this value.

\subsubsection*{Microphone}
Two Nondirectional microphones (M30, Earthworks) were attached to the end effector and the top of the 3D printer, respectively, using a PLA plastic mount.
The microphone attached to the end-effector, referred to as the main microphone, captures the sound generated when the rubber tip rubs against the textured surface.
The microphone mounted on the top of the 3D printer, referred to as the sub-microphone, captures the mechanical noise of the 3D printer for noise cancellation purposes.
The frequency response of the microphone is extremely flat, from 10 Hz to 30 kHz, with a sensitivity of 39.5 mV/Pa.
Two microphone outputs were connected to a PC via a USB audio interface (UR22mkII, Steinberg).
The sampling rate of the recordings was 44.1 kHz.
We recorded the timestamp at the start of microphone recording and used it for time synchronization with other sensors.

\subsubsection*{Load Cell}
A load cell (SC616C-1kg, Sensor and Control) measures the contact force of the rubber tip through a spring mechanism.
The load cell has a measuring range of 0.005 N to 9.8 N and an accuracy of 0.005 N.
The load cell output was connected to an A/D converter (HX711, Avia Semiconductor Ltd) which transmitted the data to the microcontroller at 80 Hz with 23-bit resolution.
The microcontroller transmitted the force values to the PC at 6 kHz via a USB serial interface.

\subsubsection*{Accelerometer}
We mounted a 3-axis MEMS accelerometer (IIS3DWB, STMicroelectronics) on a rubber tip to record acceleration information.
This accelerometer was also used in the previous study~\cite{10.1109/TOH.2024.3356609}, with a measurement range of \textpm 16 g and an accuracy of 75 µg/√Hz.
The accelerometer was installed to align the coordinate axes of the 3D printer and the accelerometer.
The sensor output was connected to a PC via a microcontroller like the load cell.
The sampling rate of the sensor values was about 6 kHz.

\subsection*{Data Preprocessing}

We implemented a data acquisition program in Python to receive sensor values on a PC.
We created separate threads for the microphone (USB audio interface), the force sensor/accelerometer (microcontroller), and the 3-axis machine for multi-threaded processing.
Each sensor value was also timestamped using the kernel clock in the operating system.

We also captured close-up color images of all textured materials using a digital image scanner (GTX830, EPSON).
The texture images stored in the dataset are square images with 1181 pixels on each side.
We cropped the texture images to a 100 mm $\times$ 100 mm region to match the probe's scanning area. The scan resolution was set to 300 dpi, resulting in a physical scale of 12 pixels/mm.

\subsection*{Textures}
\begin{figure*}[t]
    \centering
    \includegraphics[width=\linewidth]{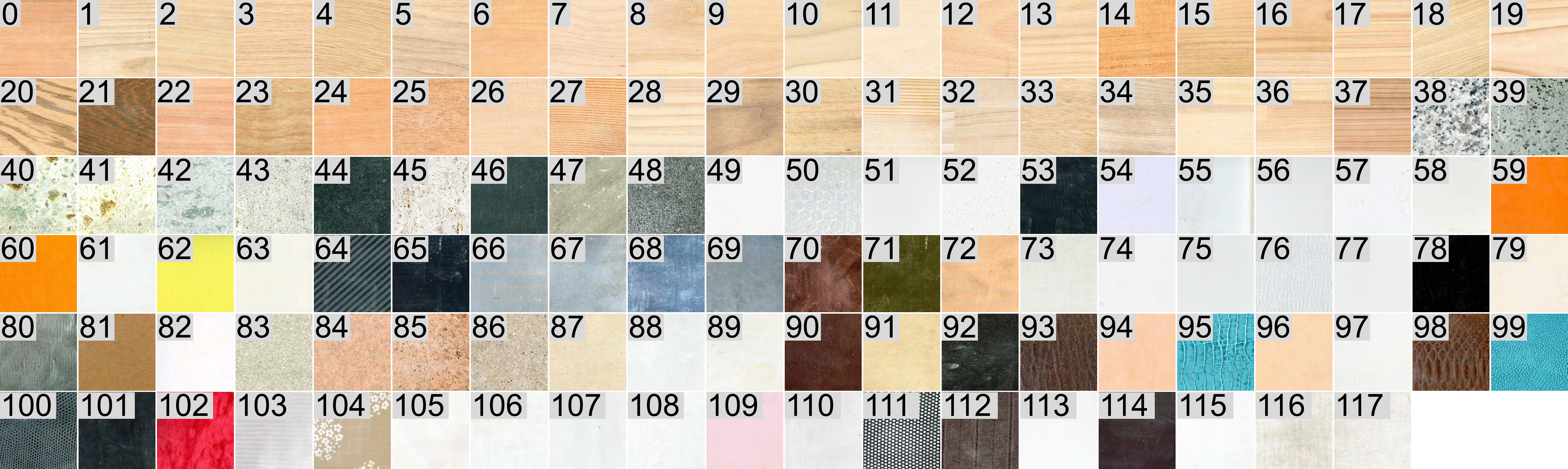}
    \caption{An overview of the images of the 118 materials that comprise the Cluster Haptic Texture Dataset. The texture materials are divided into 9 categories (0-37: Wood, 38-48: Stone, 49-64: Polymer, 65-73 Metal, 74-78: Glass, 79-81: Composite, 82-86: Ceramic, 87-101: Biological Leather, 102-117: Cloth).}
    \label{fig:textures}
\end{figure*}
Based on the texture materials from the LMT Haptic Database proposed in~\cite{Strese2014-ye}, we have collected 118 materials in 9 categories, as shown in Fig.~\ref{fig:textures}.
The materials include 38 types of wood, 11 types of stone, 16 types of polymer, 9 types of metal, 5 types of glass, 3 types of composite, 5 types of ceramic, 15 types of biological leather, and 16 types of clothes.
Each material was used as purchased, without any additional surface processing, and therefore exhibits a wide range of surface conditions, from smooth to rough textures.
Detailed information about all materials is summarized in the Supplementary information and documentation files within the our dataset on figshare~\cite{Eguchi2025texture_figshare}.

We attached these materials with adhesive to a thin magnetic sheet (Mag sheet, Nichie Co., Ltd.) to secure them to the metal bed of the 3-axis machine.

\subsection*{Recording Procedure}
Prior to haptic data recording, we captured all texture images with their positions and orientations carefully controlled to match the 100 mm $\times$ 100 mm traced area using a flatbed scanner. 

We subsequently recorded four signals for each trial using the recording system: (i) audio generated at contact, (ii) rubber tip acceleration, (iii) normal force, and (iv) the in-plane $(x,y)$ position of the rubber tip.
We tested five velocities (20, 30, 40, 50, and 60 mm/s), eight directions (intercardinal directions), and two applied forces (0.5 and 1.0 N).
The selection of sliding velocities and applied forces was based on typical ranges observed in finger–object sliding interactions~\cite{Isleyen2020-wl, Vardar2017-ct, Meyer2014-xo}.

The recording proceeded as follows for each of the 118 textures:
\begin{enumerate}
    \item Place the material on the bed of the equipment.
        \item For each force level (0.5 and 1.0 N), determine the contact height that yields the desired normal force and set this height as the local reference ($z{=}0$) for all passes at that force level. The initial position (center of the bed) is set to $x = 0$, $y = 0$.
        \item For each combination of eight intercardinal directions and five velocity conditions, plan a straight-line trajectory of length 80~mm from the origin along the specified heading, and record sensor values from the start to the end point of each sliding. Repeat this procedure twice for each condition. In total, 8 (directions) $\times$ 5 (velocities) $\times$ 2 (applied forces) $\times$ 2 (repetitions) = 160 sliding motions are performed for each material.
\end{enumerate}

After data recording, we performed post-processing including active noise cancellation of audio to remove mechanical noise and temporal synchronization between sensors. The details of active noise cancellation are described in the next Section.

\subsection*{Active Noise Cancellation System} \label{subsec:noise_cancellation}

\begin{figure}[t]
    \centering
    \includegraphics[width=0.7\linewidth]{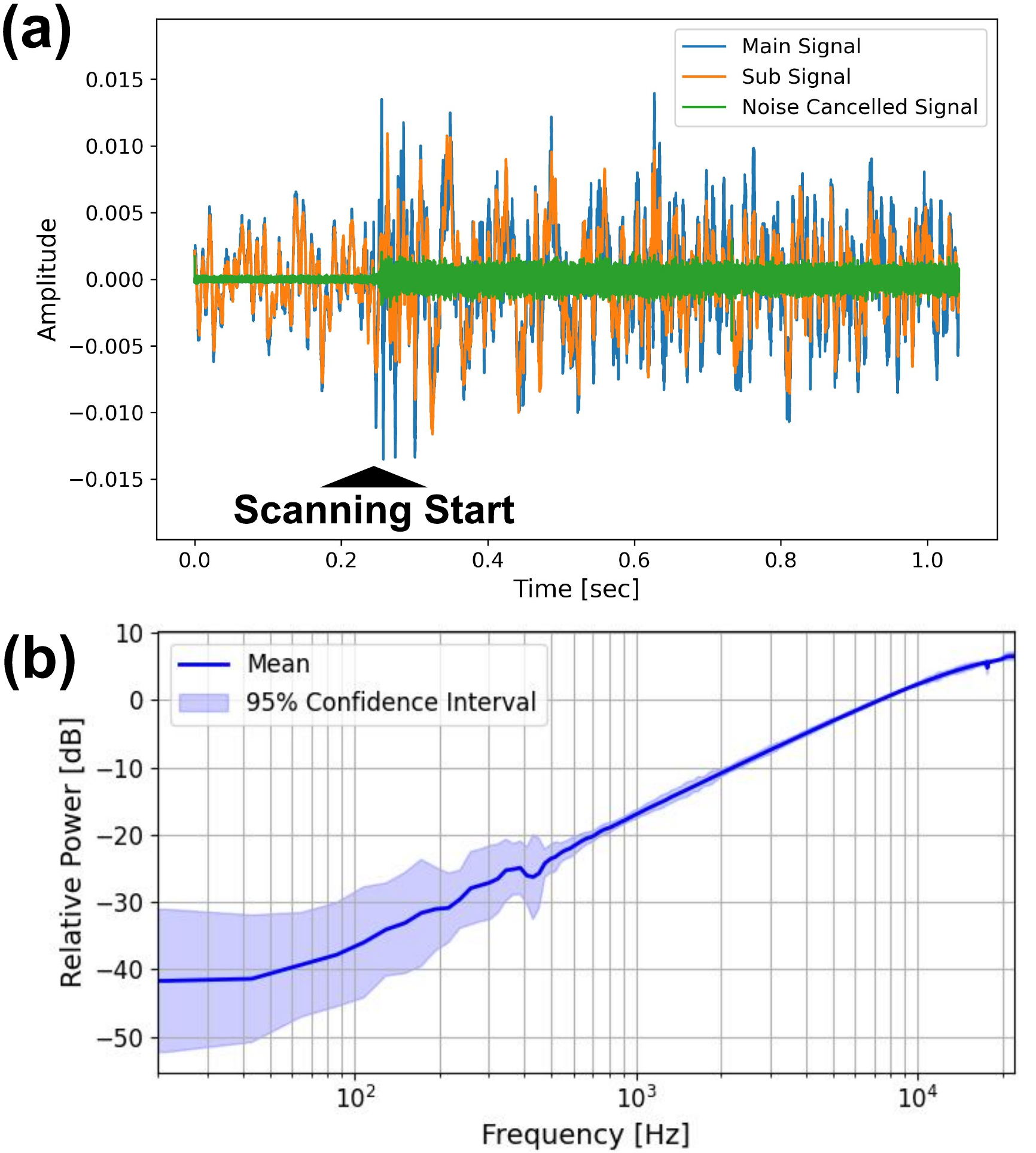}
    \caption{Active noise cancellation performance. (a) Microphone signals before and after noise cancellation. (b) Transfer function of noise cancellation applied to all sound data, showing 20-40 dB attenuation in the low-frequency range below 1000 Hz where mechanical noise is expected.}
    \label{fig:noise_cancellation}
\end{figure}

We implemented an active noise cancellation system using two microphone signals to reduce the mechanical noise generated by the 3D printer during texture scanning.
The system reduces noise by subtracting the mechanical noise recorded by a sub microphone from the main microphone signal that contains both texture-induced sounds and mechanical noise.
We employed the Normalized Least Mean Squares (NLMS) algorithm \cite{Stearns1985-qe} as a noise cancellation method. The NLMS algorithm is an adaptive Finite Impulse Response (FIR) filtering technique that iteratively updates its coefficients with a normalized step size so as to minimize the instantaneous squared error between the two microphone signal.
The system cancels noise in post processing, after the microphone signals have been recorded.

Fig.~\ref{fig:noise_cancellation}(a) shows the microphone signals before and after noise cancellation. The main signal contains both the texture-induced sounds and mechanical noise, while the noise cancelled signal shows significant reduction in noise components.
Fig.~\ref{fig:noise_cancellation}(b) shows the transfer function of our noise cancellation system. As illustrated in the figure, our implementation provides stronger noise attenuation characteristics in the lower frequency range. Specifically, in the low-frequency region below 1000 Hz where mechanical noise is predominant, we achieved noise reduction of 20-40 dB.

\subsection*{Friction Measurement}
\begin{figure}
    \centering
    \includegraphics[width=1\linewidth]{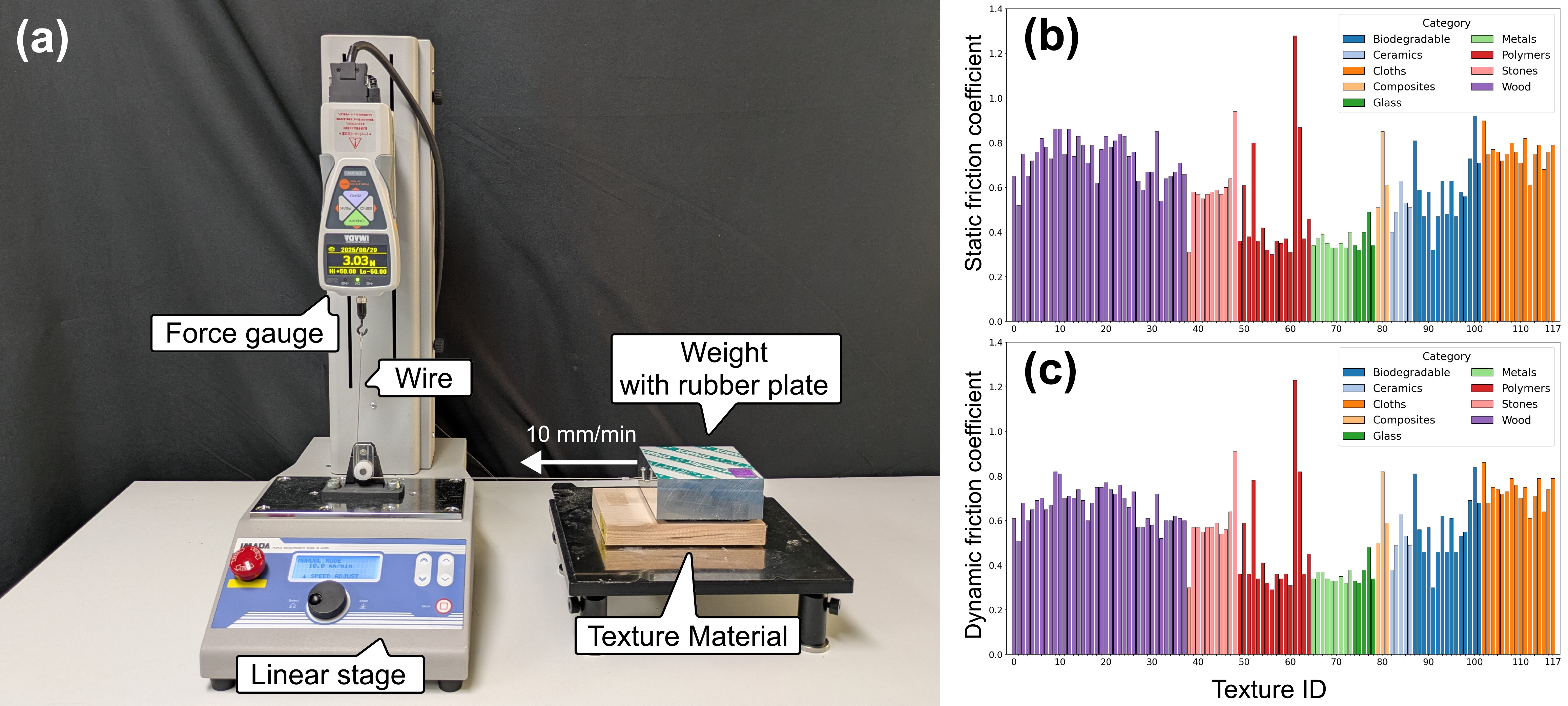}
    \caption{Overview of the friction measuring device. (a) A force gauge measures the force required to slide a weight with a urethane-rubber surface  (the same material as the rubber tip) across textured samples at a constant speed of 10 mm/min. (b)(c) Static and dynamic friction coefficients of all texture materials included in the dataset.}
    \label{fig:friction_measurement}
\end{figure}

In this dataset, we measured friction information for 118 types of materials using a device as shown in Fig.~\ref{fig:friction_measurement}.
This device consists of a linear stage (MX2-500N, IMADA), a force gauge(ZTS-50N, IMADA), a weight, and a urethane rubber sheet (same material as the rubber tip).
The linear stage moves the force gauge horizontally, which pulls the weight through a connecting wire at a constant velocity of 10 mm/min across the texture material.
The force gauge measures the force required to move the weight. For all materials, we calculated the static and dynamic friction coefficients as the ratio of measured force to the weight’s load.
We determined the static friction coefficient using the peak force at the moment the weight started moving, and the kinetic friction coefficient using the average frictional force over the 10 s following the onset of motion.
The measured friction coefficients for all texture materials are described in the documentation files included with the dataset.

\section*{Data Records}

The Cluster Haptic Texture Dataset is publicly available on figshare~\cite{Eguchi2025texture_figshare}.
The dataset is detailed in the following section.

% Directory structure is described as a hierarchical list
\subsection*{Directory Structure}
The directory structure of the Cluster Haptic Texture Dataset is as follows:

\begin{itemize}
    \item \texttt{texture\_dataset/}
    \begin{itemize}
        \item \texttt{images/}
        \begin{itemize}
            \item \texttt{images\_scan\_area/}
            \item \texttt{crop\_images/}
        \end{itemize}
        \item \texttt{sensor\_data/}
        \begin{itemize}
            \item \texttt{audio/}
            \item \texttt{raw\_audio/}
            \item \texttt{accel/}
            \item \texttt{force/}
            \item \texttt{position/}
        \end{itemize}
        \item \texttt{texture\_list.xlsx}
        \item README.MD
    \end{itemize}
\end{itemize}

At the top level, the \texttt{texture\_dataset/} directory contains two main subdirectories: \texttt{images/} for image data and \texttt{sensor\_data/} for sensor recordings, as well as two documentation files: \texttt{texture\_list.xlsx} and \texttt{README.MD}.

Within the \texttt{images/} directory, there are two subfolders: \texttt{images\_scan\_area/}, which contains images of the 100~$\times$~100~mm area scanned by the rubber tip, and \texttt{crop\_images/}, which contains 10 randomly cropped 224$\times$224 pixel images per texture, generated from \texttt{images\_scan\_area/} for use in image-based experiments.

The \texttt{sensor\_data/} directory is further divided into subfolders for each sensor modality: \texttt{audio/} for noise-canceled audio data, \texttt{raw\_audio/} for the original, unprocessed audio recordings, \texttt{accel/} for acceleration datas, \texttt{force/} for force measurements, and \texttt{position/} for position datas.
Each of these directories contains subdirectories named 0, 1, ..., 117, where each number corresponds to a texture ID (see Fig.~\ref{fig:textures}).

\texttt{texture\_list.xlsx} lists all 118 textures with their IDs, names, categories and friction coefficients. 
\texttt{README.MD} explains the directory structure, file naming and sensor informations.

\subsection*{File Naming Convention}
All sensor data files follow the naming convention: 

\texttt{(texture\_id)\_(direction)\_(velocity)\_(force)\_(repeat\_count).wav} or \texttt{.csv}. The \texttt{.wav} files are for audio data, while the \texttt{.csv} files are for acceleration and force data.

The naming convention for sensor data files is as follows:
\begin{itemize}
    \item \textbf{texture\_id}: Texture material identifier (0-117)
    \item \textbf{direction}: Clockwise angle in degrees (0, 45, 90, 135, 180, 225, 270, 315 degree)
    \item \textbf{velocity}: Scan speed in mm/s (20, 30, 40, 50, 60 mm/s)
    \item \textbf{force}: Push force in mN (500 or 1000 mN)
    \item \textbf{repeat\_count}: Number of scan repetitions (0 or 1)
\end{itemize}

Example: \texttt{0\_135\_40\_500\_0.csv} represents texture ID 0, scanned at 135 degrees, 40 mm/s velocity, 500 mN force, and repeat count 0.

\subsection*{Data Details}

The \texttt{images\_scan\_area/} directory contains images of the 100~$\times$~100~mm area scanned by the rubber tip. These images are stored in JPEG format with 300 DPI resolution, with an image size of 1181$\times$1181 pixels and a physical size of 100 mm $\times$ 100 mm.

The \texttt{crop\_images/} directory contains 10 randomly cropped 224$\times$224 pixel images per texture, with the cropping positions chosen randomly from the \texttt{images\_scan\_area/}. These images are also stored in JPEG format with 300 DPI resolution.

The \texttt{audio/} directory contains noise-canceled audio data stored in WAV format as mono signals with a sampling rate of 44.1 kHz.

The \texttt{raw\_audio/} directory contains the original, unprocessed audio recordings stored in WAV format as stereo signals with a sampling rate of 44.1 kHz. Channel 0 contains the main microphone signal (texture + machine noise), while Channel 1 contains the sub microphone signal (machine noise only).

The \texttt{accel/} directory contains acceleration data stored in CSV format with columns for time (s), X-axis acceleration (g), Y-axis acceleration (g), and Z-axis acceleration (g), recorded at a sampling rate of 6 kHz.

The \texttt{force/} directory contains force data stored in CSV format with columns for time (s) and force (N), recorded at a sampling rate of 6 kHz.

The \texttt{position/} directory contains position data stored in CSV format with columns for time (s), X-axis position (mm), and Y-axis position (mm), recorded at a sampling rate of 100 Hz.

\section*{Technical Validation}

We aimed to verify whether our dataset's sensor data, particularly audio and acceleration information, contains sufficient distinctive features for individual identification of textures, velocities, and directions.
We conducted three classification tasks: texture, velocity, and direction classification. 
In Experiment I, we performed texture classification as a preliminary experiment to identify the best classifier and verify dataset features.
In Experiment II, we examined how velocity and direction inputs affect texture classification accuracy. 
Finally, in Experiments III and IV, we examined how classification accuracy varies across different textures to investigate whether the features for each texture contain sufficient information regarding velocity and direction.

\subsection*{Setup}
\subsubsection*{Data Preprocessing}
We first downsampled the sound data to 22.05 kHz and the acceleration data to 5 kHz to reduce computational cost while preserving relevant information. After downsampling, we segmented both signals into 1-second intervals and converted each segment into a log-mel spectrogram. The log-mel spectrogram provides a time-frequency representation on a perceptually motivated logarithmic scale, which approximates the human auditory system's frequency resolution and yields rich and discriminative features for machine learning tasks involving temporal signals~\cite{gong21b_interspeech}.
For the sound data, we computed log-mel spectrograms using an FFT window size of 2048, hop length of 512, and 128 Mel bands. For the acceleration data, we used an FFT window size of 512, hop length of 128, and 128 Mel bands.

We resized all spectrograms to 224 $\times$ 224 pixels to match the input requirements of the classification models. During model training, we applied data augmentation techniques such as horizontal and vertical flipping and random erasing to improve model robustness.

\subsubsection*{Classifiers}

\begin{figure}[t]
    \centering
    \includegraphics[width=0.7\linewidth]{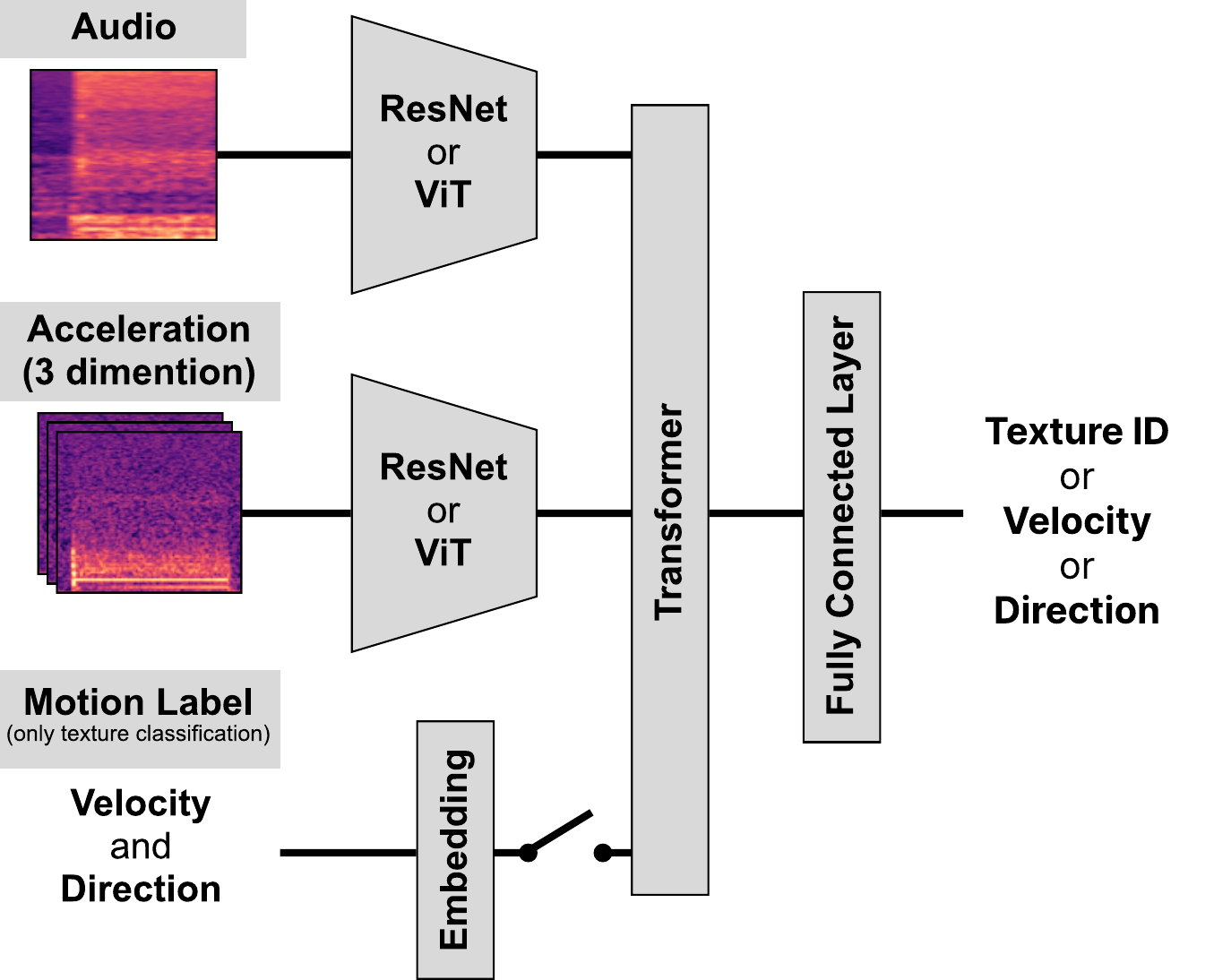}
    \caption{Architecture of our multimodal classifier. The network processes audio and acceleration signals through separate encoders (ResNet/ViT), fuses them via transformer layers, and outputs classification results through fully connected layers. For texture classification in Experiment II, motion labels (velocity/direction) are embedded into the transformer layers.}
    \label{fig:multimodal_architecture}
\end{figure}

For the classifiers, we employed both neural network-based models and classical machine learning algorithms to process our dataset's sensor data and output classification labels for texture ID, velocity, and direction based on the specific task requirements.
For neural network-based approaches, we employed CNN (ResNet)~\cite{He2015-vy} and Vision Transformer (ViT)~\cite{dosovitskiy2020image}. These models require two-dimensional array data as input, which aligns well with our log-mel spectrogram representations. We constructed these neural network classifiers by applying transfer learning from pre-trained models to our tactile texture dataset. 
For classical machine learning approaches, we utilized Support Vector Classification (SVC), decision tree, and random forest algorithms. These classifiers have been successfully used in previous haptic signal classification studies~\cite{Rong2016-ki, Devillard2023-in}. We compared the performance of all these classifiers to determine the most effective approach for texture classification.

% multimodal
We also constructed a multimodal classifier that combines different types of sensor data in our dataset.Fig.~\ref{fig:multimodal_architecture} illustrates our model architecture. 
We first extract features from audio and acceleration signals using separate CNN (ResNet) or ViT-based encoders.
These modality-specific features are then fused through transformer layers that learn cross-modal interactions through self-attention mechanisms, which enable the model to capture complex relationships between different modalities and enhance the representation of multimodal features~\cite{10.1109/TPAMI.2023.3275156, li2023see}.
Finally, the fused representations are passed through fully connected layers to produce the classification output.
For the multimodal model, we set the output dimension of each modality-specific encoder and embedding layer to 256. The transformer layers process these 256-dimensional features from both modalities, and the final output dimension before the fully connected layers is also maintained at 256.

For texture classification in Experiment II, we added velocity and direction labels as inputs. These motion labels were processed through an MLP to create embeddings, which were then integrated into the transformer layers with the audio and acceleration features. This approach helps the model learn how different velocities and directions affect the sensory signals, potentially improving classification accuracy.
For classical machine learning approaches, we concatenated the log-mel spectrograms from audio and acceleration signals along the feature dimension, following the previous multimodal sensing study~\cite{Devillard2023-in}.

\subsubsection*{Training and Evaluation}
We trained our models using the Adam optimizer with a learning rate of 1e-4. The training process utilized a batch size of 32 and ran for 50 epochs. We implemented a step learning rate scheduler with a step size of 7 epochs and a decay factor of 0.1, and applied label smoothing of 0.1.
In the experiments, we divided the data from our texture dataset in the ratio of 70:10:20 for training, validation, and testing, respectively, to train and evaluate the classifiers.
Accuracy, Precision, Recall, and F1 Score were used as indices for the evaluation of the classifiers.
We performed five training and classification cycles for each classifier with different random seeds and calculated the average and standard deviation of the evaluation indices.
We conducted all our network experiments on a machine equipped with 36 GB RAM, an Intel Core i7-12650H CPU, and a GeForce RTX 4060 Laptop with 8GB.

\subsection*{Experiment I: Texture Classification}
We performed a texture classification task on 118 different texture types using the various sensor data recorded in the Cluster Haptic Texture Dataset in this experiment.
\subsubsection*{Classification using audio signals}

\begin{table}[t]
    \centering
    \caption{Performance of classifiers using audio signal data.}
    \label{tab:audio_classification}
    \scalebox{0.84}{
        \begin{tabular}{c|c c c c}
            \textbf{Classifiers} & \textbf{Accuracy (\%)} & \textbf{Precision (\%)} & \textbf{Recall (\%)} & \textbf{F1 Score (\%)} \\
            \hline
            CNN (ResNet) & 92.05 \textpm 0.25 & 92.12 \textpm 0.18 & 92.07 \textpm 0.24 & 91.95 \textpm 0.21 \\
            ViT & 91.95 \textpm 0.69 & 92.08 \textpm 0.61 & 92.06 \textpm 0.71 & 91.89 \textpm 0.64 \\
            SVC & 63.73 \textpm 0.35 & 65.84 \textpm 0.38 & 64.00 \textpm 0.43 & 63.39 \textpm 0.31 \\
            Random Forest & 39.61 \textpm 0.46 & 40.80 \textpm 0.61 & 40.09 \textpm 0.49 & 38.76 \textpm 0.36 \\
            Decision Tree & 13.12 \textpm 0.45 & 13.53 \textpm 0.53 & 13.11 \textpm 0.44 & 13.13 \textpm 0.44 \\
        \end{tabular}
    }
\end{table}

We first performed a classification task using only the audio data.
The performances of each classifier are summarized in Table~\ref{tab:audio_classification}.
Both neural network-based approaches (CNN and ViT) achieved approximately 92\% accuracy. In contrast, classical machine learning models performed significantly worse, with the best performer (SVC) only reaching about 64\% accuracy.

\subsubsection*{Classification using acceleration signals}
\begin{table}[t]
    \centering
    \caption{Performance of classifiers using acceleration signal data.}
    \label{tab:accel_classification}
    \scalebox{0.84}{
        \begin{tabular}{c|c c c c}
            \textbf{Classifiers} & \textbf{Accuracy (\%)} & \textbf{Precision (\%)} & \textbf{Recall (\%)} & \textbf{F1 Score (\%)} \\
            \hline
            CNN (ResNet) & 69.38 \textpm 2.18 & 69.79 \textpm 2.20 & 69.61 \textpm 2.31 & 69.11 \textpm 2.34 \\
            ViT & 56.31 \textpm 2.55 & 58.35 \textpm 2.70 & 56.43 \textpm 2.47 & 56.05 \textpm 2.69 \\
            SVC & 15.75 \textpm 0.56 & 22.21 \textpm 2.43 & 16.21 \textpm 0.49 & 15.36 \textpm 0.56 \\
            Random Forest & 20.14 \textpm 0.48 & 20.30 \textpm 0.71 & 20.16 \textpm 0.42 & 19.42 \textpm 0.53 \\
            Decision Tree & 7.80 \textpm 0.51 & 8.01 \textpm 0.49 & 7.78 \textpm 0.64 & 7.76 \textpm 0.55 \\
        \end{tabular}
    }
\end{table}
We also performed a classification task using only the acceleration data.
The performances of each classifier are shown in Table~\ref{tab:accel_classification}.
Similar to audio classification, neural network-based approaches outperformed classical machine learning models. However, even the best-performing CNN classifier achieved only about 69\% accuracy, which is significantly lower than the results obtained with audio signals. Additionally, the ViT model showed approximately 10\% lower accuracy compared to the CNN model.

\subsubsection*{Classification using audio and acceleration signals}
\begin{table}[t]
    \centering
    \caption{Performance of classifiers using both audio and acceleration signal data.}
    \label{tab:audio_accel_classification}
    \scalebox{0.84}{
        \begin{tabular}{c|c c c c}
            \textbf{Classifiers} & \textbf{Accuracy (\%)} & \textbf{Precision (\%)} & \textbf{Recall (\%)} & \textbf{F1 Score (\%)} \\
            \hline
            CNN (Resnet) & 95.95 \textpm 2.08 & 96.16 \textpm 1.89 & 96.04 \textpm 2.12 & 95.96 \textpm 2.12 \\
            ViT & 87.83 \textpm 0.35 & 88.77 \textpm 0.31 & 87.93 \textpm 0.31 & 87.76 \textpm 0.30 \\
            SVC & 32.84 \textpm 0.47 & 37.18 \textpm 1.09 & 33.10 \textpm 0.36 & 32.37 \textpm 0.53 \\
            Random Forest & 33.87 \textpm 0.54 & 34.36 \textpm 0.72 & 34.40 \textpm 0.61 & 32.82 \textpm 0.55 \\
            Decision Tree & 13.37 \textpm 0.63 & 13.78 \textpm 0.53 & 13.28 \textpm 0.67 & 13.33 \textpm 0.62 \\
        \end{tabular}
    }
\end{table}

We investigated the classification performance of a multimodal classifier combining audio and acceleration data.
The performance of each classifier are shown in Table~\ref{tab:audio_accel_classification}.
It was found that training with multimodal information improved the accuracy of the CNN-based classifier to approximately 96\%, which is higher than when using single-modal information. However, for the other classification methods, the accuracy actually deteriorated compared to using audio signals alone.

\subsubsection*{Summary}
\begin{figure}[!t]
    \centering
    \includegraphics[width=1.0\columnwidth]{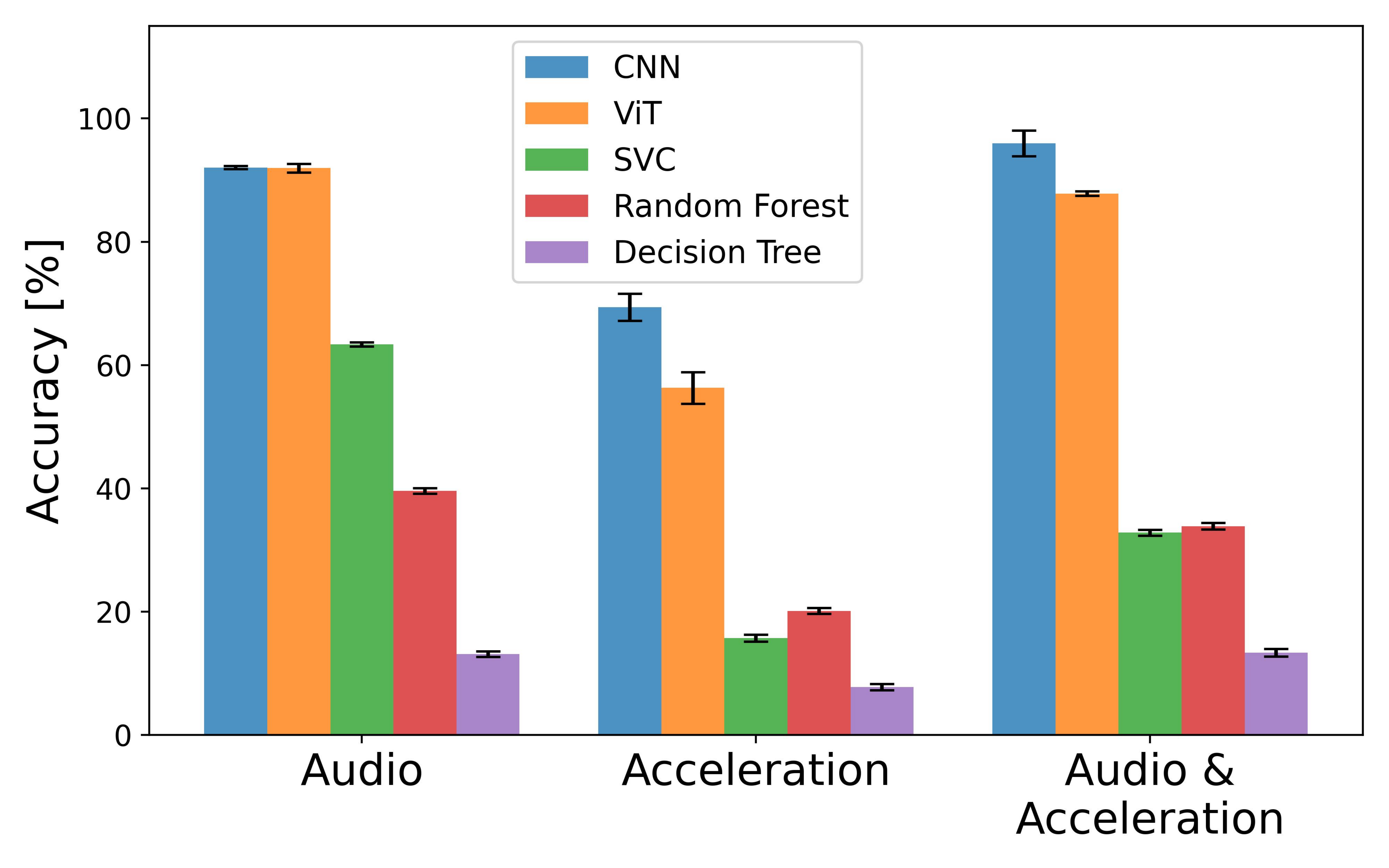}
    \caption{Classification accuracies for each condition in the texture classifier using five machine learning models. The values represent the average of five training and classification cycles. Error bars indicate standard deviations.}
    \label{fig:texture_classification_summary}
\end{figure}
Figure~\ref{fig:texture_classification_summary} shows the classification results for each condition.
The CNN-based classifier demonstrated significantly superior accuracy across all modalities compared to the other machine-learning models. 
Particularly, the CNN classifier achieved approximately 96\% accuracy when using multimodal data (audio and acceleration).
These findings confirm that the 118 textures in our dataset exhibit unique, classifiable features that can be effectively captured by high-capacity models such as CNNs.
The significant performance gap between audio-based and acceleration-based classifiers suggests that acceleration data may not capture texture-specific features as effectively as audio signals. This lower performance with acceleration data may be due to the inherently less distinctive features in acceleration signals compared to frictional sounds. Alternatively, it is possible that the device developed in this study did not sufficiently capture the relevant characteristics of the acceleration signals.

\begin{figure}[!t]
    \centering
    \includegraphics[width=1.0\columnwidth]{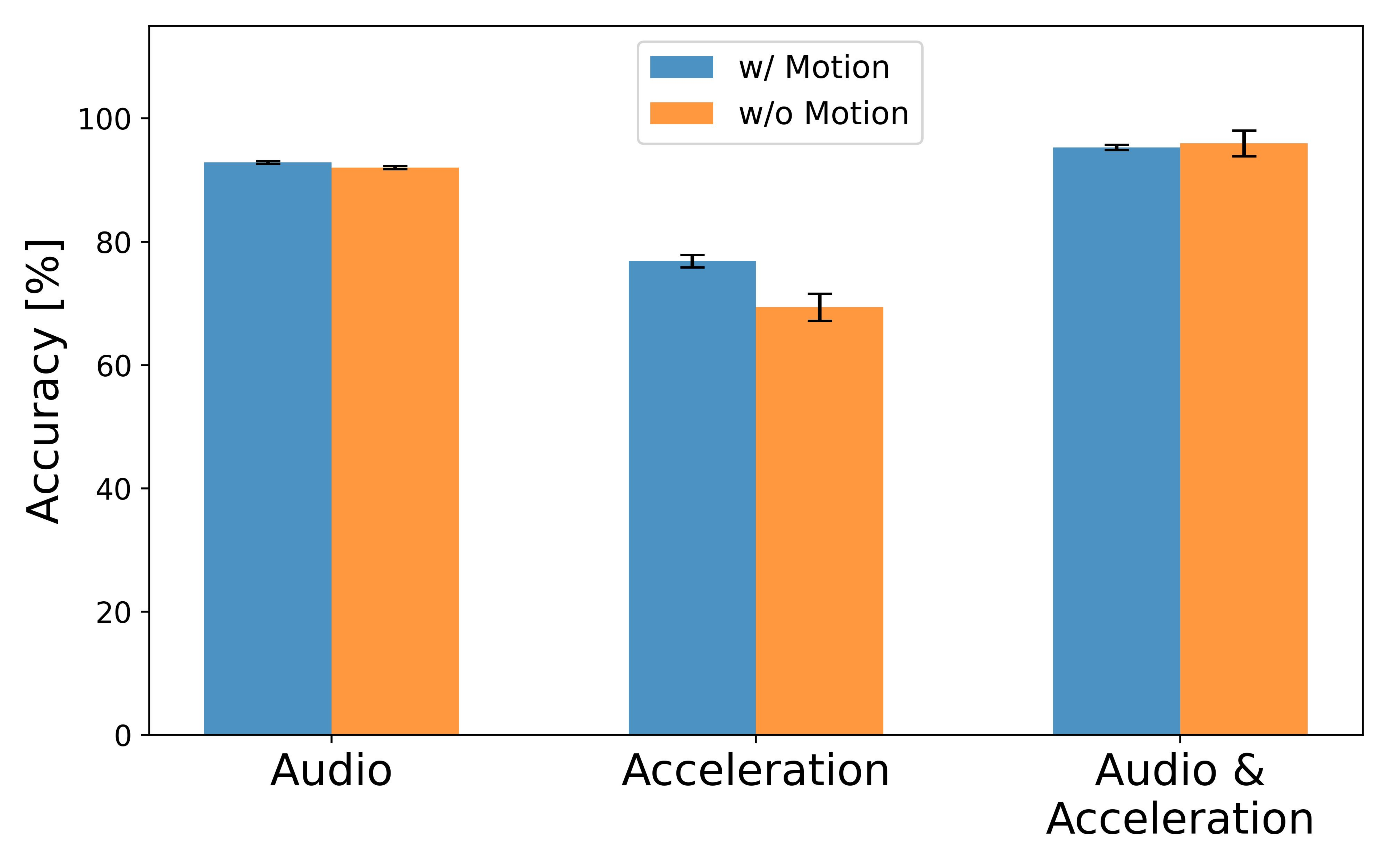}
    \caption{Classification accuracies of classifiers using audio and acceleration signal data with and without motion labels. Error bars indicate standard deviations.}
    \label{fig:motion_classification}
\end{figure}
\begin{table}[!t]
    \centering
    \caption{Classification accuracies(\%) of classifiers using audio and acceleration signal data with and without motion labels.}
    \label{tab:motion_classification}
        \begin{tabular}{c|ccc}
        & Audio & Acceleration & Audio \& Acceleration \\
        \hline
        w/ Motion   & 92.85 \textpm 0.22 & 76.88 \textpm 1.01 & 95.31 \textpm 0.42 \\
        w/o Motion  & 92.05 \textpm 0.25 & 69.38 \textpm 2.18 & 95.95 \textpm 2.08 \\
        \end{tabular}
\end{table}

\subsection*{Experiment II: Texture Classification with Motion Labels}
In this experiment, as a practical use case of our dataset, we evaluated how integrating motion information (velocity and direction) as additional features with haptic signals such as audio and acceleration affects classification accuracy. 

We used the CNN (ResNet-34) model, which performed best in Experiment I, as our baseline and conducted texture classification with the multimodal architecture shown in Fig.~\ref{fig:multimodal_architecture}. The classifier was trained using audio, acceleration, and combined signals, both with and without motion labels (velocity and direction), to examine the impact of motion information on classification accuracy.

Table~\ref{tab:motion_classification} and Fig.~\ref{fig:motion_classification} summarize the classification accuracies of classifiers using audio and acceleration signal data with and without motion labels.
The results showed that adding motion labels to both audio and acceleration signals led to a slight improvement in classification accuracy. Notably, the acceleration-based classifier showed the most significant improvement, with an increase of approximately 7\% in accuracy. However, for the combined audio and acceleration signals, no improvement in accuracy was observed when motion labels were added. 

% The improvement in classification accuracy when adding motion labels to acceleration signals suggests that explicit motion information helps the classifier focus on texture-specific features in the acceleration data. This phenomenon might be analogous to how humans integrate motion cues in texture perception, where motion parameters can influence the perception of surface properties. 
% However, the limited improvement in accuracy for audio-based and combined audio-acceleration classifiers indicates that audio signals already contain sufficient texture-specific information, making additional motion cues less beneficial. This could be due to the inherent stability of audio signals across different motion parameters, which may not significantly alter the texture-specific characteristics captured by the audio modality. 
% Furthermore, the high sampling rate of audio data might capture features beyond human perceptual capabilities, suggesting that the relationship between motion and texture perception in our system may differ from human perception.

\subsection*{Experiment III: Velocity Classification}

\begin{figure*}[t]
    \centering
    \includegraphics[width=1.0\linewidth]{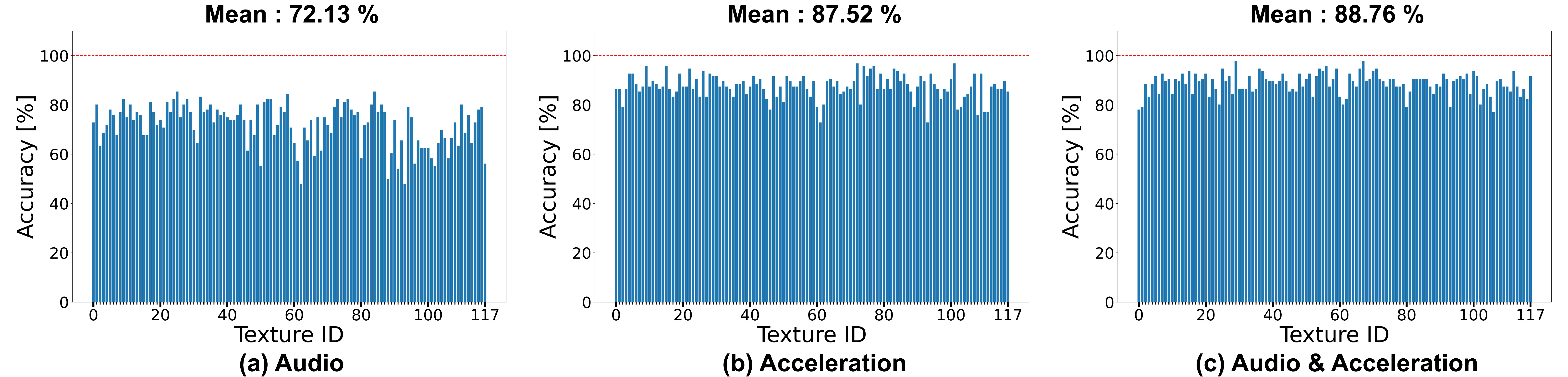}
    \caption{Accuracy per texture for CNN-based classifiers in velocity classification. The classifiers were trained on (a) audio signals, (b) acceleration signals, and (c) combined audio and acceleration signals, respectively. The accuracy is higher for classifiers trained with audio signals compared to those trained with acceleration signals.}
    \label{fig:velocity_classification}
\end{figure*}
\begin{table}[!t]
    \centering
    \caption{Top 5 textures with high or low accuracies in velocity label classification using both audio and acceleration signals.}
    \label{tab:velocity_classification}
    \begin{tabular}{lc|lc}
        \multicolumn{2}{c|}{Texture w/ high accuracy} & \multicolumn{2}{c}{Texture w/ low accuracy} \\ \hline
        \multicolumn{1}{c}{ID: Name}   & Accuracy (\%)  & \multicolumn{1}{c}{ID: Name}  & Accuracy (\%) \\ \hline
        29: Walnut & 97.92   & 106: Cotton gauze & 77.08   \\
        67: Cast iron  & 97.92 & 93: Goat skin & 78.12   \\
        56: PTFE & 95.83   & 0: Nyatoh & 79.16   \\
        36: Hinoki & 94.79   & 1: Castor aralia & 80.20   \\
        66: Steel & 94.79 & 80: Tar paper & 80.20  
    \end{tabular}
\end{table}
\begin{figure}[!t]
    \centering
    \includegraphics[width=\columnwidth]{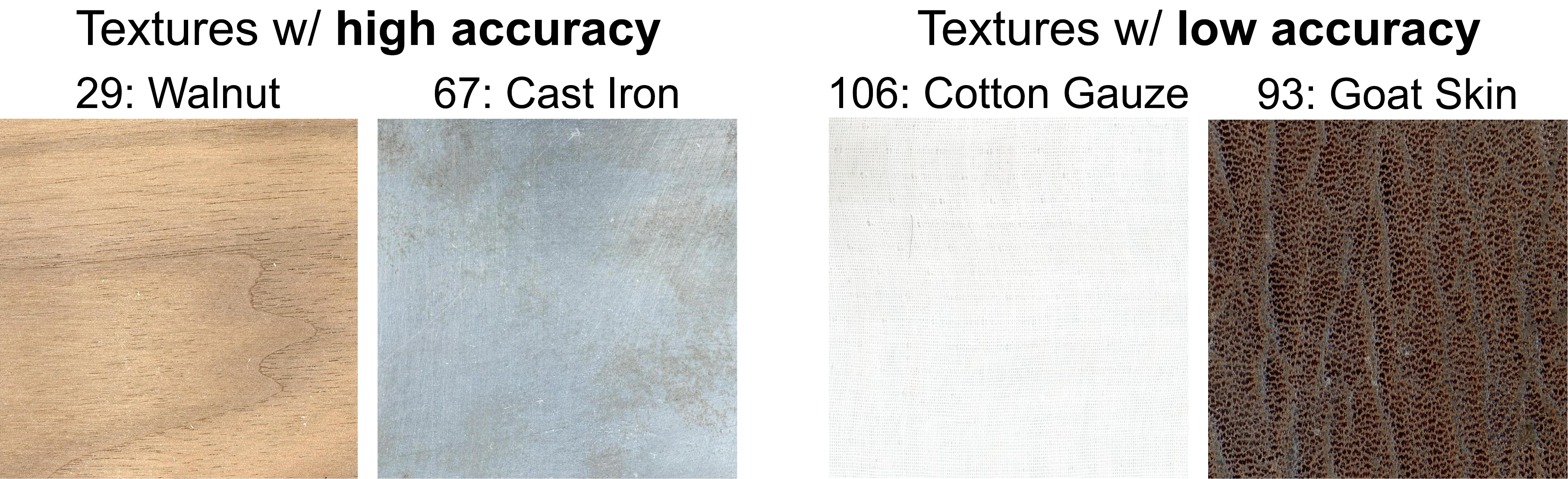}
    \caption{Images of Top 2 textures showing high and low accuracies in velocity label classification using both audio and acceleration signals.}
    \label{fig:texture_diff_velocity}
\end{figure}

We classified five velocities to evaluate how variations in the velocity of tracing affect the sensor signal in our haptic texture dataset.

We used a CNN-based classifier for velocity classification, as it showed the best performance in Experiment I. In this experiment, five velocities were used as labels instead of texture IDs. For each texture, we trained a separate classifier and calculated the velocity classification accuracy to examine the influence of texture properties.

Fig.~\ref{fig:velocity_classification}a-c shows the velocity label classification accuracies by texture for each classifier.
The accuracies are highest when both audio and acceleration signals are used, similar to Experiment I, with an average accuracy of 88.76~\%.
% Interestingly, unlike the texture classification results in Experiment I, classifiers using only acceleration signals outperformed those using only audio signals in the velocity classification task.

A list of texture material types with high and low accuracies and their accuracies for classifiers using both audio and acceleration signals is shown in Table~\ref{tab:velocity_classification}.
Table~\ref{tab:velocity_classification} shows that walnut and cast iron textures (Fig.~\ref{fig:texture_diff_velocity}a) have relatively high accuracy.
These textures have very smooth surfaces and low coefficients of friction, so the sensor signal measurements change regularly with changes in velocity, and the velocity labels can be classified with high accuracy.

On the other hand, materials with low accuracies, such as cotton gauze and goat skin (Fig.~\ref{fig:texture_diff_velocity}b), possess challenging physical properties for velocity classification. Cotton gauze is thin and soft with a high coefficient of friction against the rubber tip, while goat skin features a notably rough and irregular surface. These materials' non-uniform structures create inconsistent mechanical interactions during traversal at different velocities. Consequently, the sensor signals exhibit unpredictable patterns that do not correlate systematically with velocity changes, making it difficult for the classifier to identify distinctive velocity-dependent features in the data.

% Nevertheless, this dataset was collected with an artificial fingertip, and the above results may not translate perfectly to interactions involving a real human finger. Even with this limitation, the combined audio-acceleration model still achieves a mean classification accuracy of 88.76 \%, substantially exceeding the chance level, which indicates that the key signal features are captured robustly by the artificial fingertip. Consequently, supplementing this dataset with recordings obtained from actual human fingertips~\cite{devillard2025tactile} should facilitate analyses and learning schemes that more closely mirror human tactile perception.

\subsection*{Experiment IV: Direction Classification}

\begin{figure*}[!t]
    \centering
    \includegraphics[width=1.0\linewidth]{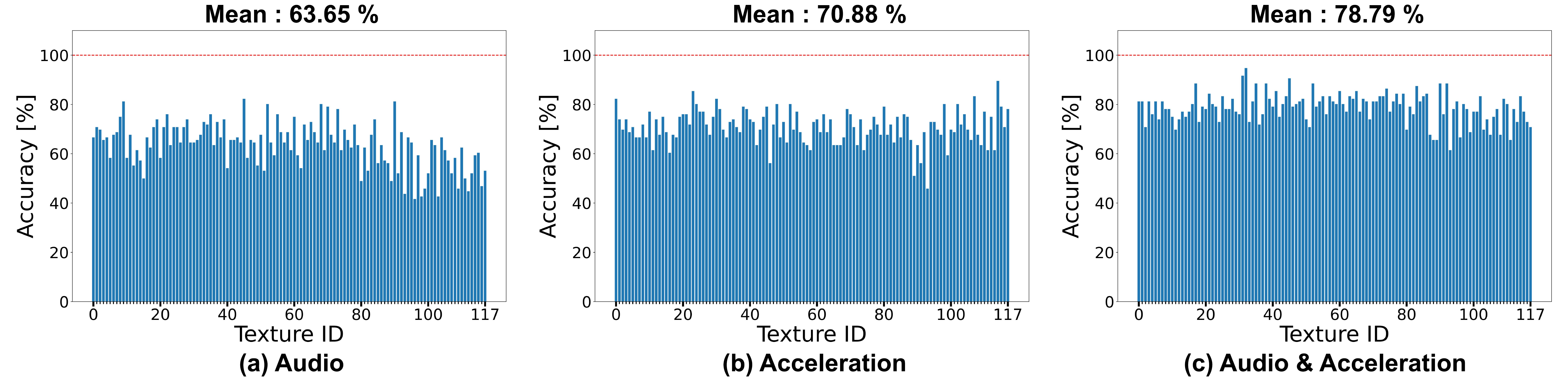}
    \caption{Accuracies per texture for CNN-based classifiers in direction classification. The classifiers were trained on (a) audio signals, (b) acceleration signals, and (c) combined audio and acceleration signals, respectively. The classifier trained with combined audio and acceleration signals shows the highest accuracy.}
    \label{fig:direction_classification}
\end{figure*}
\begin{table}[!t]
    \centering
    \caption{Top 5 textures with high or low accuracies in direction label classification using both audio and acceleration signals.}
    \label{tab:direction_classification}
    \begin{tabular}{lc|lc}
        \multicolumn{2}{c|}{Texture w/ high accuracy} & \multicolumn{2}{c}{Texture w/ low accuracy} \\ \hline
        \multicolumn{1}{c}{ID: Name}   & Accuracy (\%) & \multicolumn{1}{c}{ID: Name}  & Accuracy (\%)  \\ \hline
        32: Gumwood & 94.79   & 93: Goat skin & 61.46   \\
        31: North pipe & 91.67 & 88: Cotton & 65.63   \\
        45: Travertine & 90.63   & 89: Wool felt & 65.63   \\
        92: Cow fur & 88.54  & 111: Mesh & 65.63   \\
        38: Granite & 88.54 & 96: Cow skin & 66.67  
    \end{tabular}
\end{table}
\begin{figure}[!t]
    \centering
    \includegraphics[width=\columnwidth]{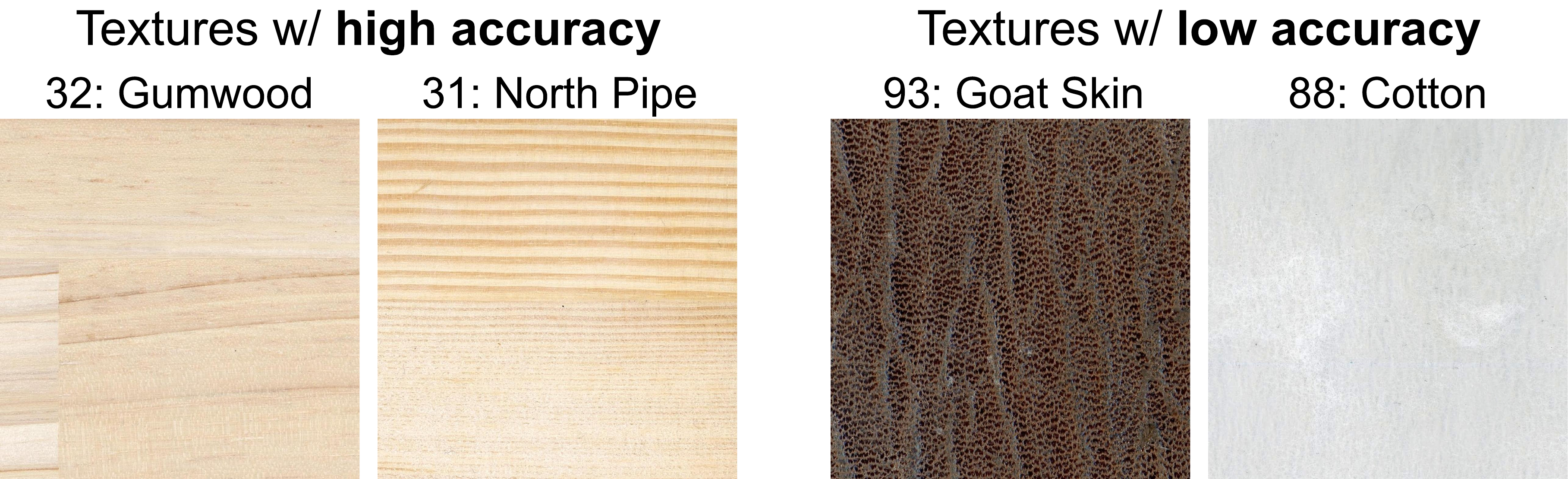}
    \caption{Images of Top 2 textures showing high and low accuracies in direction label classification using both audio and acceleration signals.}
    \label{fig:texture_diff_direction}
\end{figure}

To clarify how changes in sliding direction affect the signal captured by the sensor in our tactile texture dataset, we performed a classification task to classify eight different sliding directions.

For direction classification, we used the same CNN-based classifier as in Experiment III. Eight traversing directions were set as labels, and classification accuracy was computed for each texture.

Fig.~\ref{fig:direction_classification}a-c shows the direction label classification accuracies by texture for each classifier.
Compared to the velocity label classification in Experiment III, the accuracy tended to be lower for all classifiers using any data in the direction label classification.
The classifier with the highest average accuracy was the one that used both audio and acceleration signals, with an average accuracy of 78.79~\%.
% The classification accuracy using the acceleration signal was higher than the audio-based model, similar to the results of Experiment III.

A list of texture material types with high and low accuracy and their 
accuracy for classifiers using both audio and acceleration signals is 
shown in Table~\ref{tab:direction_classification}. Notably, the textures with the highest accuracy were gumwood and north pipe (Fig.~\ref{fig:texture_diff_direction}a).
These woods exhibit pronounced, quasi-periodic wood grain structures along a dominant orientation. When the rubber tip traverses the surface, the number of wood grain ridges crossed per unit time changes systematically with scan direction, producing direction-specific stick-slip events reflected in the acceleration and acoustic spectra. 

Conversely, materials with low accuracy, such as goat skin and cotton (Fig.~\ref{fig:texture_diff_direction}b), lack dominant periodic patterns, and their low elastic moduli allow substantial surface deformation, dissipating high-frequency energy. The resulting acceleration and audio signals exhibit broadband, noise-like spectra that are highly similar across directions, making reliable classification difficult.

\section*{Code Availability}
The dataset used in this study is publicly available on figshare~\cite{Eguchi2025texture_figshare}. The experimental code is available at: 

\url{https://github.com/cluster-lab/Cluster-Haptic-Texture-Dataset}.

The accompanying README.md file provides detailed instructions for dataset installation and usage of the experimental code.

% =============================================================
%                        BACK MATTER
% =============================================================

% ---- References ----
% \bibliographystyle{naturemag}
% \bibliography{references}

\begin{thebibliography}{10}
\expandafter\ifx\csname url\endcsname\relax
  \def\url#1{\texttt{#1}}\fi
\expandafter\ifx\csname urlprefix\endcsname\relax\def\urlprefix{URL }\fi
\providecommand{\bibinfo}[2]{#2}
\providecommand{\eprint}[2][]{\url{#2}}

\bibitem{Lederman2009-tf}
\bibinfo{author}{Lederman, S.~J.} \& \bibinfo{author}{Klatzky, R.~L.}
\newblock \bibinfo{title}{Haptic perception: A tutorial}.
\newblock \emph{\bibinfo{journal}{Attention, Perception \& Psychophysics}} \textbf{\bibinfo{volume}{71}}, \bibinfo{pages}{1439--1459} (\bibinfo{year}{2009}).
\newblock \urlprefix\url{https://doi.org/10.3758/APP.71.7.1439}.

\bibitem{ryan2021interaction}
\bibinfo{author}{Ryan, C.~P.} \emph{et~al.}
\newblock \bibinfo{title}{The interaction between motion and texture in the sense of touch}.
\newblock \emph{\bibinfo{journal}{Journal of Neurophysiology}} \textbf{\bibinfo{volume}{126}}, \bibinfo{pages}{1375--1390} (\bibinfo{year}{2021}).
\newblock \urlprefix\url{https://doi.org/10.1152/jn.00583.2020}.

\bibitem{roberts2024visual}
\bibinfo{author}{Roberts, R.~D.}, \bibinfo{author}{Li, M.} \& \bibinfo{author}{Allen, H.~A.}
\newblock \bibinfo{title}{Visual effects on tactile texture perception}.
\newblock \emph{\bibinfo{journal}{Scientific Reports}} \textbf{\bibinfo{volume}{14}}, \bibinfo{pages}{632} (\bibinfo{year}{2024}).
\newblock \urlprefix\url{https://doi.org/10.1038/s41598-023-50596-1}.

\bibitem{El_Saddik2007-wl}
\bibinfo{author}{El~Saddik, A.}
\newblock \bibinfo{title}{The potential of haptics technologies}.
\newblock \emph{\bibinfo{journal}{IEEE Instrumentation \& Measurement Magazine}} \textbf{\bibinfo{volume}{10}}, \bibinfo{pages}{10--17} (\bibinfo{year}{2007}).
\newblock \urlprefix\url{https://doi.org/10.1109/MIM.2007.339540}.

\bibitem{Bau2010-nk}
\bibinfo{author}{Bau, O.}, \bibinfo{author}{Poupyrev, I.}, \bibinfo{author}{Israr, A.} \& \bibinfo{author}{Harrison, C.}
\newblock \bibinfo{title}{{TeslaTouch}: electrovibration for touch surfaces}.
\newblock In \emph{\bibinfo{booktitle}{Proceedings of the 23rd Annual {ACM} Symposium on User Interface Software and Technology ({UIST})}}, \bibinfo{pages}{283--292} (\bibinfo{year}{2010}).
\newblock \urlprefix\url{https://doi.org/10.1145/1866029.1866074}.

\bibitem{Miyatake2023-bo}
\bibinfo{author}{Miyatake, Y.}, \bibinfo{author}{Hiraki, T.}, \bibinfo{author}{Iwai, D.} \& \bibinfo{author}{Sato, K.}
\newblock \bibinfo{title}{{HaptoMapping}: Visuo-haptic augmented reality by embedding user-imperceptible tactile display control signals in a projected image}.
\newblock \emph{\bibinfo{journal}{IEEE Transactions on Visualization and Computer Graphics}} \textbf{\bibinfo{volume}{29}}, \bibinfo{pages}{2005--2019} (\bibinfo{year}{2023}).
\newblock \urlprefix\url{https://doi.org/10.1109/TVCG.2021.3136214}.

\bibitem{Ito2022-vr}
\bibinfo{author}{Ito, M.}, \bibinfo{author}{Sakuma, R.}, \bibinfo{author}{Ishizuka, H.} \& \bibinfo{author}{Hiraki, T.}
\newblock \bibinfo{title}{{AirHaptics}: Vibrotactile presentation method using an airflow from audio speakers of smart devices}.
\newblock In \emph{\bibinfo{booktitle}{Proceedings of the 28th {ACM} Symposium on Virtual Reality Software and Technology ({VRST})}}, \bibinfo{number}{Article 39}, \bibinfo{pages}{1--2} (\bibinfo{year}{2022}).
\newblock \urlprefix\url{https://doi.org/10.1145/3562939.3565670}.

\bibitem{Wiertlewski2011-es}
\bibinfo{author}{Wiertlewski, M.}, \bibinfo{author}{Hudin, C.} \& \bibinfo{author}{Hayward, V.}
\newblock \bibinfo{title}{On the 1/f noise and non-integer harmonic decay of the interaction of a finger sliding on flat and sinusoidal surfaces}.
\newblock In \emph{\bibinfo{booktitle}{Proceedings of 2011 {IEEE} World Haptics Conference ({WHC})}}, \bibinfo{pages}{25--30} (\bibinfo{year}{2011}).
\newblock \urlprefix\url{https://doi.org/10.1109/WHC.2011.5945456}.

\bibitem{Platkiewicz2014-th}
\bibinfo{author}{Platkiewicz, J.}, \bibinfo{author}{Mansutti, A.}, \bibinfo{author}{Bordegoni, M.} \& \bibinfo{author}{Hayward, V.}
\newblock \bibinfo{title}{Recording device for natural haptic textures felt with the bare fingertip}.
\newblock In \emph{\bibinfo{booktitle}{Haptics: Neuroscience, Devices, Modeling, and Applications (Proceedings of 2014 {EuroHaptics} Conference)}}, \bibinfo{pages}{521--528} (\bibinfo{year}{2014}).
\newblock \urlprefix\url{https://doi.org/10.1007/978-3-662-44193-0_65}.

\bibitem{8793967}
\bibinfo{author}{Taunyazov, T.}, \bibinfo{author}{Koh, H.~F.}, \bibinfo{author}{Wu, Y.}, \bibinfo{author}{Cai, C.} \& \bibinfo{author}{Soh, H.}
\newblock \bibinfo{title}{Towards effective tactile identification of textures using a hybrid touch approach}.
\newblock In \emph{\bibinfo{booktitle}{2019 International Conference on Robotics and Automation (ICRA)}}, \bibinfo{pages}{4269--4275} (\bibinfo{year}{2019}).
\newblock \urlprefix\url{https://doi.org/10.1109/ICRA.2019.8793967}.

\bibitem{10.1145/3586183.3606764}
\bibinfo{author}{Kitagishi, T.}, \bibinfo{author}{Hiroi, Y.}, \bibinfo{author}{Watanabe, Y.}, \bibinfo{author}{Itoh, Y.} \& \bibinfo{author}{Rekimoto, J.}
\newblock \bibinfo{title}{Telextiles: End-to-end remote transmission of fabric tactile sensation}.
\newblock In \emph{\bibinfo{booktitle}{Proceedings of the 36th Annual ACM Symposium on User Interface Software and Technology (UIST)}} (\bibinfo{year}{2023}).
\newblock \urlprefix\url{https://doi.org/10.1145/3586183.3606764}.

\bibitem{huang2021texture}
\bibinfo{author}{Huang, S.} \& \bibinfo{author}{Wu, H.}
\newblock \bibinfo{title}{Texture recognition based on perception data from a bionic tactile sensor}.
\newblock \emph{\bibinfo{journal}{Sensors}} \textbf{\bibinfo{volume}{21}}, \bibinfo{pages}{5224} (\bibinfo{year}{2021}).
\newblock \urlprefix\url{https://doi.org/10.3390/s21155224}.

\bibitem{Culbertson2013-ut}
\bibinfo{author}{Culbertson, H.}, \bibinfo{author}{Unwin, J.}, \bibinfo{author}{Goodman, B.~E.} \& \bibinfo{author}{Kuchenbecker, K.~J.}
\newblock \bibinfo{title}{Generating haptic texture models from unconstrained tool-surface interactions}.
\newblock In \emph{\bibinfo{booktitle}{Proceedings of 2013 World Haptics Conference ({WHC})}}, \bibinfo{pages}{295--300} (\bibinfo{year}{2013}).
\newblock \urlprefix\url{https://doi.org/10.1109/WHC.2013.6548424}.

\bibitem{10.1109/TOH.2024.3521418}
\bibinfo{author}{Hosoi, J.}, \bibinfo{author}{Jin, D.}, \bibinfo{author}{Ban, Y.} \& \bibinfo{author}{Warisawa, S.}
\newblock \bibinfo{title}{Voluminous fur stroking experience through interactive visuo-haptic model in virtual reality}.
\newblock \emph{\bibinfo{journal}{IEEE Trans. Haptics}} \textbf{\bibinfo{volume}{18}}, \bibinfo{pages}{232–243} (\bibinfo{year}{2024}).
\newblock \urlprefix\url{https://doi.org/10.1109/TOH.2024.3521418}.

\bibitem{Deng2009-uz}
\bibinfo{author}{Deng, J.} \emph{et~al.}
\newblock \bibinfo{title}{Imagenet: A large-scale hierarchical image database}.
\newblock In \emph{\bibinfo{booktitle}{Proceedings of 2009 {IEEE} Conference on Computer Vision and Pattern Recognition ({CVPR})}}, \bibinfo{pages}{248--255} (\bibinfo{year}{2009}).
\newblock \urlprefix\url{https://doi.org/10.1109/CVPR.2009.5206848}.

\bibitem{7952261}
\bibinfo{author}{Gemmeke, J.~F.} \emph{et~al.}
\newblock \bibinfo{title}{Audio set: An ontology and human-labeled dataset for audio events}.
\newblock In \emph{\bibinfo{booktitle}{2017 IEEE International Conference on Acoustics, Speech and Signal Processing (ICASSP)}}, \bibinfo{pages}{776--780} (\bibinfo{year}{2017}).
\newblock \urlprefix\url{https://doi.org/10.1109/ICASSP.2017.7952261}.

\bibitem{Culbertson2014-go}
\bibinfo{author}{Culbertson, H.}, \bibinfo{author}{Lopez~Delgado, J.~J.} \& \bibinfo{author}{Kuchenbecker, K.~J.}
\newblock \bibinfo{title}{One hundred data-driven haptic texture models and open-source methods for rendering on {3D} objects}.
\newblock In \emph{\bibinfo{booktitle}{Proceedings of 2014 {IEEE} Haptics Symposium ({HAPTICS})}}, \bibinfo{pages}{319--325} (\bibinfo{year}{2014}).
\newblock \urlprefix\url{https://doi.org/10.1109/HAPTICS.2014.6775475}.

\bibitem{Zheng2016-ge}
\bibinfo{author}{Zheng, H.} \emph{et~al.}
\newblock \bibinfo{title}{Deep learning for surface material classification using haptic and visual information}.
\newblock \emph{\bibinfo{journal}{IEEE Transactions on Multimedia}} \textbf{\bibinfo{volume}{18}}, \bibinfo{pages}{2407--2416} (\bibinfo{year}{2016}).
\newblock \urlprefix\url{https://doi.org/10.1109/TMM.2016.2598140}.

\bibitem{Strese2014-ye}
\bibinfo{author}{Strese, M.} \emph{et~al.}
\newblock \bibinfo{title}{A haptic texture database for tool-mediated texture recognition and classification}.
\newblock In \emph{\bibinfo{booktitle}{Proceedings of 2014 {IEEE} International Symposium on Haptic, Audio and Visual Environments and Games ({HAVE})}}, \bibinfo{pages}{118--123} (\bibinfo{year}{2014}).
\newblock \urlprefix\url{https://doi.org/10.1109/HAVE.2014.6954342}.

\bibitem{strese2016multimodal}
\bibinfo{author}{Strese, M.}, \bibinfo{author}{Schuwerk, C.}, \bibinfo{author}{Iepure, A.} \& \bibinfo{author}{Steinbach, E.}
\newblock \bibinfo{title}{Multimodal feature-based surface material classification}.
\newblock \emph{\bibinfo{journal}{IEEE transactions on haptics}} \textbf{\bibinfo{volume}{10}}, \bibinfo{pages}{226--239} (\bibinfo{year}{2016}).
\newblock \urlprefix\url{https://doi.org/10.1109/TOH.2016.2625787}.

\bibitem{7989057}
\bibinfo{author}{Burka, A.}, \bibinfo{author}{Rajvanshi, A.}, \bibinfo{author}{Allen, S.} \& \bibinfo{author}{Kuchenbecker, K.~J.}
\newblock \bibinfo{title}{Proton 2: Increasing the sensitivity and portability of a visuo-haptic surface interaction recorder}.
\newblock In \emph{\bibinfo{booktitle}{2017 IEEE International Conference on Robotics and Automation (ICRA)}}, \bibinfo{pages}{439--445} (\bibinfo{year}{2017}).
\newblock \urlprefix\url{https://doi.org/10.1109/ICRA.2017.7989057}.

\bibitem{10.1109/TOH.2024.3356609}
\bibinfo{author}{Khojasteh, B.}, \bibinfo{author}{Shao, Y.} \& \bibinfo{author}{Kuchenbecker, K.~J.}
\newblock \bibinfo{title}{Robust surface recognition with the maximum mean discrepancy: Degrading haptic-auditory signals through bandwidth and noise}.
\newblock \emph{\bibinfo{journal}{IEEE Trans. Haptics}} \textbf{\bibinfo{volume}{17}}, \bibinfo{pages}{58--65} (\bibinfo{year}{2024}).
\newblock \urlprefix\url{https://doi.org/10.1109/TOH.2024.3356609}.

\bibitem{Jiao2019-hd}
\bibinfo{author}{Jiao, J.}, \bibinfo{author}{Zhang, Y.}, \bibinfo{author}{Wang, D.}, \bibinfo{author}{Guo, X.} \& \bibinfo{author}{Sun, X.}
\newblock \bibinfo{title}{{HapTex}: A database of fabric textures for surface tactile display}.
\newblock In \emph{\bibinfo{booktitle}{Proceedings of 2019 {IEEE} World Haptics Conference ({WHC})}}, \bibinfo{pages}{331--336} (\bibinfo{year}{2019}).
\newblock \urlprefix\url{https://doi.org/10.1109/WHC.2019.8816167}.

\bibitem{devillard2025tactile}
\bibinfo{author}{Devillard, A.~W.}, \bibinfo{author}{Ramasamy, A.}, \bibinfo{author}{Cheng, X.}, \bibinfo{author}{Faux, D.} \& \bibinfo{author}{Burdet, E.}
\newblock \bibinfo{title}{Tactile, audio, and visual dataset during bare finger interaction with textured surfaces}.
\newblock \emph{\bibinfo{journal}{Scientific Data}} \textbf{\bibinfo{volume}{12}}, \bibinfo{pages}{484} (\bibinfo{year}{2025}).
\newblock \urlprefix\url{https://doi.org/10.1038/s41597-025-04670-0}.

\bibitem{8794285}
\bibinfo{author}{Takahashi, K.} \& \bibinfo{author}{Tan, J.}
\newblock \bibinfo{title}{Deep visuo-tactile learning: Estimation of tactile properties from images}.
\newblock In \emph{\bibinfo{booktitle}{2019 International Conference on Robotics and Automation (ICRA)}}, \bibinfo{pages}{8951--8957} (\bibinfo{year}{2019}).
\newblock \urlprefix\url{https://doi.org/10.1109/ICRA.2019.8794285}.

\bibitem{Cai2022-tg}
\bibinfo{author}{Cai, S.} \emph{et~al.}
\newblock \bibinfo{title}{{GAN-based} image-to-friction generation for tactile simulation of fabric material}.
\newblock \emph{\bibinfo{journal}{Computers \& Graphics}} \textbf{\bibinfo{volume}{102}}, \bibinfo{pages}{460--473} (\bibinfo{year}{2022}).
\newblock \urlprefix\url{https://doi.org/10.1016/j.cag.2021.09.007}.

\bibitem{Song2023-yv}
\bibinfo{author}{Song, R.}, \bibinfo{author}{Sun, X.} \& \bibinfo{author}{Liu, G.}
\newblock \bibinfo{title}{{Cross-Modal} generation of tactile friction coefficient from audio and visual measurements by transformer}.
\newblock \emph{\bibinfo{journal}{IEEE Transactions on Instrumentation and Measurement}}  (\bibinfo{year}{2023}).
\newblock \urlprefix\url{https://doi.org/10.1109/TIM.2023.3311071}.

\bibitem{liu2024maniwav}
\bibinfo{author}{Liu, Z.} \emph{et~al.}
\newblock \bibinfo{title}{Maniwav: Learning robot manipulation from in-the-wild audio-visual data}.
\newblock In \emph{\bibinfo{booktitle}{Proceedings of the 2024 Conference on Robot Learning ({CoRL})}}, \bibinfo{pages}{947--962} (\bibinfo{year}{2024}).

\bibitem{li2023see}
\bibinfo{author}{Li, H.} \emph{et~al.}
\newblock \bibinfo{title}{See, hear, and feel: Smart sensory fusion for robotic manipulation}.
\newblock In \emph{\bibinfo{booktitle}{In Proceedings of the 2022 Conference on Robot Learning ({CoRL})}}, \bibinfo{pages}{1368--1378} (\bibinfo{year}{2022}).

\bibitem{Kuramitsu2013-os}
\bibinfo{author}{Kuramitsu, K.}, \bibinfo{author}{Nomura, T.}, \bibinfo{author}{Nomura, S.}, \bibinfo{author}{Maeno, T.} \& \bibinfo{author}{Nonomura, Y.}
\newblock \bibinfo{title}{Friction evaluation system with a human finger model}.
\newblock \emph{\bibinfo{journal}{Chemistry Letters}} \textbf{\bibinfo{volume}{42}}, \bibinfo{pages}{284--285} (\bibinfo{year}{2013}).
\newblock \urlprefix\url{https://doi.org/10.1246/cl.2013.284}.

\bibitem{Kouchi2005-pq}
\bibinfo{author}{Kouchi, M.}, \bibinfo{author}{Miyata, N.} \& \bibinfo{author}{Mochimaru, M.}
\newblock \bibinfo{title}{An analysis of hand measurements for obtaining representative japanese hand models}.
\newblock In \emph{\bibinfo{booktitle}{{SAE} Technical Paper}}, \bibinfo{number}{2005-01-2734}, \bibinfo{pages}{1--7} (\bibinfo{year}{2005}).
\newblock \urlprefix\url{https://doi.org/10.4271/2005-01-2734}.

\bibitem{Artificial_Intelligence_Research_Center_AIST_undated-jc}
\bibinfo{author}{{Artificial Intelligence Research Center, AIST}}.
\newblock \bibinfo{title}{{AIST} japanese hand dimension data}.
\newblock \bibinfo{howpublished}{\url{https://www.airc.aist.go.jp/dhrt/hand/data/list.html}}.
\newblock \bibinfo{note}{Accessed: 2023-9-25}.

\bibitem{Eguchi2025texture_figshare}
\bibinfo{author}{Eguchi, M.}, \bibinfo{author}{Hayase, T.}, \bibinfo{author}{Hiroi, Y.} \& \bibinfo{author}{Hiraki, T.}
\newblock \bibinfo{title}{Cluster haptic texture dataset: Haptic texture dataset with variety in velocity and direction of sliding contacts} (\bibinfo{year}{2025}).
\newblock \urlprefix\url{https://doi.org/10.6084/m9.figshare.29438288}.

\bibitem{Isleyen2020-wl}
\bibinfo{author}{Isleyen, A.}, \bibinfo{author}{Vardar, Y.} \& \bibinfo{author}{Basdogan, C.}
\newblock \bibinfo{title}{Tactile roughness perception of virtual gratings by electrovibration}.
\newblock \emph{\bibinfo{journal}{IEEE Transactions on Haptics}} \textbf{\bibinfo{volume}{13}}, \bibinfo{pages}{562--570} (\bibinfo{year}{2020}).
\newblock \urlprefix\url{https://doi.org/10.1109/TOH.2019.2959993}.

\bibitem{Vardar2017-ct}
\bibinfo{author}{Vardar, Y.}, \bibinfo{author}{Isleyen, A.}, \bibinfo{author}{Saleem, M.~K.} \& \bibinfo{author}{Basdogan, C.}
\newblock \bibinfo{title}{Roughness perception of virtual textures displayed by electrovibration on touch screens}.
\newblock In \emph{\bibinfo{booktitle}{Proceedings of 2017 {IEEE} World Haptics Conference ({WHC})}}, \bibinfo{pages}{263--268} (\bibinfo{year}{2017}).
\newblock \urlprefix\url{https://doi.org/10.1109/WHC.2017.7989912}.

\bibitem{Meyer2014-xo}
\bibinfo{author}{Meyer, D.~J.}, \bibinfo{author}{Wiertlewski, M.}, \bibinfo{author}{Peshkin, M.~A.} \& \bibinfo{author}{Colgate, J.~E.}
\newblock \bibinfo{title}{Dynamics of ultrasonic and electrostatic friction modulation for rendering texture on haptic surfaces}.
\newblock In \emph{\bibinfo{booktitle}{Proceedings of 2014 {IEEE} Haptics Symposium ({HAPTICS})}}, \bibinfo{pages}{63--67} (\bibinfo{year}{2014}).
\newblock \urlprefix\url{https://doi.org/10.1109/HAPTICS.2014.6775434}.

\bibitem{Stearns1985-qe}
\bibinfo{author}{Stearns, B.~W.} \& \bibinfo{author}{S}.
\newblock \bibinfo{title}{Adaptive signal processing}.
\newblock \emph{\bibinfo{journal}{Prentice Hall}}  (\bibinfo{year}{1985}).

\bibitem{gong21b_interspeech}
\bibinfo{author}{Gong, Y.}, \bibinfo{author}{Chung, Y.-A.} \& \bibinfo{author}{Glass, J.}
\newblock \bibinfo{title}{Ast: Audio spectrogram transformer}.
\newblock In \emph{\bibinfo{booktitle}{Interspeech 2021}}, \bibinfo{pages}{571--575} (\bibinfo{year}{2021}).
\newblock \urlprefix\url{10.21437/Interspeech.2021-698}.

\bibitem{He2015-vy}
\bibinfo{author}{He, K.}, \bibinfo{author}{Zhang, X.}, \bibinfo{author}{Ren, S.} \& \bibinfo{author}{Sun, J.}
\newblock \bibinfo{title}{Deep residual learning for image recognition}.
\newblock In \emph{\bibinfo{booktitle}{2016 IEEE Conference on Computer Vision and Pattern Recognition (CVPR)}}, \bibinfo{pages}{770--778} (\bibinfo{year}{2016}).
\newblock \urlprefix\url{https://doi.org/10.1109/CVPR.2016.90}.

\bibitem{dosovitskiy2020image}
\bibinfo{author}{Dosovitskiy, A.} \emph{et~al.}
\newblock \bibinfo{title}{An image is worth 16x16 words: Transformers for image recognition at scale}.
\newblock In \emph{\bibinfo{booktitle}{In Proceedings of the 2021 International Conference on Learning Representations ({ICLR})}} (\bibinfo{year}{2021}).

\bibitem{Rong2016-ki}
\bibinfo{author}{Rong, F.}
\newblock \bibinfo{title}{Audio classification method based on machine learning}.
\newblock In \emph{\bibinfo{booktitle}{Proceedings of 2016 International Conference on Intelligent Transportation, Big Data \& Smart City ({ICITBS})}}, \bibinfo{pages}{81--84} (\bibinfo{year}{2016}).
\newblock \urlprefix\url{https://doi.org/10.1109/ICITBS.2016.98}.

\bibitem{Devillard2023-in}
\bibinfo{author}{Devillard, A.}, \bibinfo{author}{Ramasamy, A.}, \bibinfo{author}{Faux, D.}, \bibinfo{author}{Hayward, V.} \& \bibinfo{author}{Burdet, E.}
\newblock \bibinfo{title}{Concurrent haptic, audio, and visual data set during bare finger interaction with textured surfaces}.
\newblock In \emph{\bibinfo{booktitle}{Proceedings of 2023 {IEEE} World Haptics Conference ({WHC})}}, \bibinfo{pages}{101--106} (\bibinfo{year}{2023}).
\newblock \urlprefix\url{https://doi.org/10.1109/WHC56415.2023.10224372}.

\bibitem{10.1109/TPAMI.2023.3275156}
\bibinfo{author}{Xu, P.}, \bibinfo{author}{Zhu, X.} \& \bibinfo{author}{Clifton, D.~A.}
\newblock \bibinfo{title}{Multimodal learning with transformers: A survey}.
\newblock \emph{\bibinfo{journal}{IEEE Transactions on Pattern Analysis and Machine Intelligence}} \textbf{\bibinfo{volume}{45}}, \bibinfo{pages}{12113–12132} (\bibinfo{year}{2023}).
\newblock \urlprefix\url{https://doi.org/10.1109/TPAMI.2023.3275156}.

\end{thebibliography}

\section*{Author Contributions}
All authors contributed to the research planning.
M.E. developed the recording system, collected data, and conducted experiments.
All authors contributed to writing and reviewing the paper and approved the final version of the content.

\section*{Competing Interests}
The authors declare no competing interests.

\section*{Acknowledgements}
This work was partially supported by JSPS KAKENHI Grant Numbers JP22H01447 and JP23H04328, Japan.

\end{document}